
\input amstex
\documentstyle{amsppt}
\magnification=1200
\NoBlackBoxes
\define\id{\operatorname{id}}
\define\Diff{\operatorname{Diff}}
\define\Rot{\operatorname{Rot}}
\define\circle{\Bbb S^1}
\define\sLtwo{\operatorname{sl}(2,\Bbb C)}
\define\sLthree{\operatorname{sl}(3,\Bbb C)}
\define\Vect{\operatorname{Vect}}
\define\CVect{\operatorname{\Bbb CVect}}
\define\vir{\operatorname{vir}}
\define\Cvir{\operatorname{\Bbb Cvir}}
\define\Vir{\operatorname{Vir}}
\define\CVir{\operatorname{\Bbb CVir}}
\define\Ang{\operatorname{Ang}}
\define\Mantle{\operatorname{Mantle}}
\define\Gf{\operatorname{Gf}}
\define\Ner{\operatorname{Ner}}
\define\LNer{\operatorname{LNer}}
\define\LDiff{\operatorname{LDiff}}
\define\pd#1{\frac{\partial}{\partial #1}}
\define\ad{\operatorname{ad}}
\redefine\Sp{\operatorname{Sp}}
\define\Const{\operatorname{Const}}
\define\Gr{\operatorname{Gr}}
\define\Hom{\operatorname{Hom}}
\define\PSl{\operatorname{PSl}}
\define\mes{\operatorname{mes}}
\redefine\span{\operatorname{span}}
\document\eightpoint
\qquad\qquad\qquad\qquad\qquad\qquad\qquad\qquad\qquad\qquad\qquad\qquad
\qquad hep-th/9403068

\

\

\bf
\centerline{INFINITE DIMENSIONAL GEOMETRY AND QUANTUM FIELD}
\centerline{THEORY OF STRINGS}
\centerline{I. INFINITE DIMENSIONAL GEOMETRY OF SECOND QUANTIZED}
\centerline{FREE STRING}
\rm
\

\

\centerline{D.JURIEV}

\

\bf This is a preliminary "manuscript" version of the text published in
the journal "Algebras, Groups and Geometries" by the independent Editor
Prof.J.L\^ohmus (Institute of Physics, Estonian Academy of Sciences, Tartu):
Alg. Groups Geom. 11 (1994) 145-179.

This publication have no any relation to the rest Editorial Board as well as
to other publications in the journal.

\

\rm

\

\

Abstract. There are investigated several objects of an infinite dimensional
geometry, appearing from the second quantization of a free string.

\

\

\

This paper is devoted to structures of an infinite dimensional geometry,
appearing in quantum field theory of (closed) strings; the objects connected
with second quantization of a free string are described in the first part of
the paper, when an analogous material for self--interacting string field is
proposed to be discussed in the second part.

In the present paper we follow the general ideology of string theory presented
in [1]. It should be mentioned that the detailed exposition of used formalism
of the first quantization of a closed string is contained in the book [2].
This publication maybe considered as a continuation of a previous one [3]
devoted to geometric aspects of quantum conformal field theory.

The first part contains two chapters: the first one is devoted to the
infinite dimensional geometry of flag, fundamental and $\Pi$--spaces for the
Virasoro--Bott group and its nonassociative deformation defined by
Gelfand--Fuchs 3--cocycle (which will be called Gelfand--Fuchs loop) as well
as of infinite--dimensional non--Euclidean symplectic grassmannian, to the
construction of Verma modules, their models and skladens over the Virasoro
algebra and their properties; in the second chapter there is described an
infinite dimensional geometry of the configuration space for the second
quantized free string in flat and curved backgrounds, as well as an author
version of Bowick--Rajeev formalism of the separation of internal and external
degrees of freedom of a closed string.

The great attention is paid to an interaction of various geometric structures:
in the first chapter they are infinite dimensional Lie algebras, groups and
loops, homogeneous, K\"ahler, Finsler, contact and symmetric spaces, complex,
real and CR--manifolds, determinant sheaves, manifolds with subsymmetries,
polarizations and Fock spaces, bibundles and objects of integral geometry,
nonholonomic spaces, deformations of geometric structures and moduli spaces;
in the second one --- gauge fields, Faddeev--Popov ghosts, Gauss--Manin
connections, Kostant--Blattner--Sternberg pairings, BRST--operators. The text
does not contain any essential terminological innnovations --- our purpose is
rather to play an interaction of known classical concepts on a nice
infinite--dimensional example.
\newpage

\head 1. Infinite dimensional geometry of the flag manifold $M(\Vir)$, the
fundamental affine space $A(\Vir)$ and $\Pi$--space $\Pi(\Vir)$ for the
virasoro--Bott group, the universal deformation of a complex disc, the
infinite dimensional symplectic non--Euclidean grassmannian $\Lambda(\Vir)$,
the spaces of angles $\Ang(\Vir)$ and punctured mirrors $\Sigma(\Vir)$ on the
flag manifold $M(\Vir)$. The Verma modules over the Virasoro algebra, their
models and skladens
\endhead

\subhead 1.1. The Lie algebra $\Vect(\circle)$ of vector fields on a circle,
the group $\Diff_+(\circle)$ of diffeomorphisms of a circle, the Virasoro
algebra $\vir$, the Virasoro--Bott group $\Vir$, the Neretin semigroup $\Ner$
(the mantle $\Mantle(\Diff_+(\circle))$ of the group of diffeomorphisms of a
circle). KIrillov construction and class $S$ of univalent functions. the
Gelfand--Fuchs 3--cocycle, Finsler geometry of $M(\Vir)$ and the
Gelfand--Fuchs loop $\Gf$, the non--associative deformation of the
Virasoro--Bott group $\Vir$
\endsubhead

Let $\Diff(\circle)=\Diff_+(\circle)\sqcup\Diff_-(\circle)$ be the group of
analytic diffeomorphisms of a circle $\circle$: diffeomorphisms from the
subgroup $\Diff_+(\circle)$ preserve an orientation on a circle $\circle$,
ones from the coset $\Diff_-(\circle)$ change it; Lie algebra of the group
$\Diff_+(\circle)$ is identified with the vector space $\Vect(\circle)$ of
analytic vector fields $v(t)d/dt$ on a circle $\circle$; the structural
constants of the complexification $\CVect(\circle)$ of the Lie algebra
$\Vect(\circle)$ have the form $c^l_{jk}=(j-k)\delta^l_{j+k}$ in the basis
$e_k=i\exp(ikt)d/dt$. In 1968 I.M.Gelfand and D.B.Fuchs discovered a
non--trivial central extension of Lie algebra $\Vect(\circle)$: the
corresponding 2--cocycle maybe written as $c(u,v)=\int u'(t)\,dv'(t)$ or as
$c(u,v)=\det(A(t_0))$, where $A(t)=\left[\matrix u'(t) & v'(t) \\ u''(t) &
v''(t)
\endmatrix\right]$; independently this central extension was discovered in
1969 by M.Virasoro and was called later the Virasoro algebra $\vir$ (the same
name belongs to the complexification $\Cvir$ of this algbera); the Virasoro
algebra $\Cvir$ is generated by the vectors $e_k$ and the central element $c$,
the commutation relations in it has the form
$[e_j,e_k]=(j-k)e_{j+k}+\delta(j+k)\cdot\frac{j^3-j}{12}\cdot c$. One may
correspond an infinite dimensional group $\Vir$ to the Lie algebra $\vir$
which is a centarl extension of the group $\Diff_+(\circle)$, the
corresponding 2--cocycle was calculated by R.Bott in 1977: $c(f,g)=\int
\log(g\circ f)'\, d\log f'$. The group $\Vir$ is called the Virasoro--Bott
group.

There are no any infinite dimensional groups corresponding to Lie algebras
$\CVect(\circle)$ and $\Cvir$ but it is useful to consider the following
construction, which is attributed to Yu.Neretin and developped by
M.Kontsevich [4,5]. Let us denote accordingly to Yu.Neretin by $\LDiff^{\Bbb
C}_+(\circle)$ a set of all analytic mappings $g:\circle\longmapsto\Bbb
C\backslash \{0\}$ with a Jordan image $g(\circle)$ such that zero belongs to
the
interior of the contour $g(\circle)$, the orientations of $g(\circle)$ and
$\circle$ coincide and the value of $g'(z)$ is not equal to zero anywhere;
$\LDiff^{\Bbb C}_+(\circle)$ is a local group in the following sense: let
$g_1$ and $g_2$ belong to $\LDiff^{\Bbb C}_+(\circle)$ and $g_1$ admits an
analytic extension to an area which contains the contour $g_2(\circle)$, then
the composition $g_1\circ g_2$ is correctly defined. let's denote by $\LNer$
the local semigroup in $\LDiff^{\Bbb C}_+(\circle)$ consisting of mappings $g$
such that $|g(\exp(it))|<1$; as it was shown by Yu.Neretin [4] local
semigroup $\LNer$ maybe supplied by a natural structure of a global semigroup
$\Ner$. There exist two different constructions of the Neretin semigroup.
{\sl The first construction} (Yu.Neretin [4,5]). an element of semigroup $\Ner$
--- a formal product $pA(t)q$ (*), where $p,q\!\in\!\Diff_+(\circle)$,
$p(1)=1$, $t>0$, $A(t):\Bbb C\mapsto \Bbb C$ such that $A(t)x=\exp(-t)z$. To
define a multiplication in $\Ner$ it is necessary to describe a rule of
transformation of formal product $A(s)pA(t)$ to the form (*). If $t$ is so
small that the diffeomorphism $p$ maybe analytically extended to the ring
$\exp(-t)\le|z|\le 1$, then there is correctly defined a product
$g=A(s)pA(t)$; let $K$ be a domain bounded by $\circle$ and $g(\circle)$, and
$Q$ is a canonical conformal mapping from $K$ onto the ring
$\exp(-t')\le|z|\le 1$ such that $Q(1)=1$, then $g=p'A(t')q'$, where $\left.
p'=Q^{-1}\right|_{\circle}$, and $q'$ is defined from the relation
$A(s)pA(t)=p'A(t')q'$. If $t$ is an arbitrary real number then there exist
$t/n$ so small that a product
$A(s)pA(t)=(\ldots((A(s)pA(t/n))A(t/n)\ldots)A(t/n)$ maybe calculated
accordingly the previous construction. the obtained multiplication is
associative [4]. {\sl The second construction} (M.Kontsevich, in the version
of Yu.Neretin). An element $g$ of semigroup $\Ner$ is a triple $(K,p,q)$,
where $K$ is a Riemann surface with a boundary, which is biholomorphically
equivalent to the ring, $p,q:\circle\longmapsto\partial K$ are fixed analytic
parametrizations of the boundary $\partial K$ of the surface $K$, so that $K$
is on the right side from $p(\exp(it))$ and from the left side from
$q(\exp(it))$. Two elements $g_1=(K_1,p_1,q_1)$ and $g_2=(K_2,p_2,q_2)$ are
equivalent if there exists a conformal mapping $R:K_1\longmapsto K_2$ such
that $p_2=Rp_1$, $q_2=Rq_1$; the product of two elements $g_1$ and $g_2$ of
the semigroup $\Ner$ is the element $g_3=(K_3,p_3,q_3)$ of this semigroup,
where $K_3=K_1\bigsqcup_{q_1(\exp(it))=p_2(\exp(it))}K_2$ and $p_3=p_2$,
$q_3=q_2$. The Neretin semigroup $\Ner$ is called the mantle of the group
$\Diff_+(\circle)$ of diffeomorphisms of a circle and is denoted by
$\Mantle(\Diff_+(\circle))$. The Neretin semigroup admits a central extension;
the corresponding 2--cocycle was calculated by Yu.Neretin in 1989 [4], the
explicit formulas for that cocycle is rather cumbersome so that they are
omitted.

The flag manifold $M(\Vir)$ for the Virasoro--Bott group $\Vir$ is the
homogeneous space $\Diff_+(\circle)/\circle$ with the group of motions
$\Diff_+(\circle)$ and the isotropy group $\circle$ [6-8]. There exist several
realizations of this manifold. The realization of $M(\Vir)$ as an infinite
dimensional homogeneous space $\Diff_+(\circle)/\circle$ is called algebraic;
in this realization the space $M(\Vir)$ maybe also identified with the
quotient of the Neretin semigroup $\Ner$ by its subsemigroup $\Ner^{\circ}$
consisting of elements $g$ of the semigroup $\Ner$, which admit an analytic
extension to $D_-$ ($D_-=\{z\!\in\!\Bbb C:|z|\ge 1\})$. The probabilistic
realization: the group $\Diff_+(\circle)$ acts on the space of all
probabilistic measures $u(t)\,dt$ on a circle with an analytic positive
density $u(t)$ in a natural way, the stabilizer of the point $(2\pi)^{-1}\,dt$
is isomorphic to $\circle$, therefore, from the transitivity of the action of
the group $\Diff_+(\circle)$ on the space of probabilistic measures it follows
that this space maybe identified with $M(\Vir)$. Orbital realization: the
space $M(\Vir)$ maybe considered as an orbit of the coadjoint representation
for the groups $\Diff_+(\circle)$ or $\Vir$ [9-11], namely, elements of the
space $\vir^*$ dual to the Virasoro algebra $\vir$ are identified with pairs
$(p(t)dt^2,b)$ so that the coadjoint action of the group $\Vir$ has the
following form $K(g)(p,b)=(gp-bS(g),b)$, where $S(g)$ is the Schwarz
derivative of a function $g$; the orbit of the point $(a\cdot dt^2,b)$
coincides with $M(\Vir)$ if and only if $a/b=-n^2/2$, $n=1, 2, 3, 4,\ldots$.
So there is defined a family of symplectic structures $\omega_{a,b}$ on the
space $M(\Vir)$. Analytic realization (Kirillov construction) [6,7]: let us
consider the space $S$ of functions $f(z)$ analytic and univalent in the
closed unit disc $D_+$ normalized by the conditions $f(0)=0$, $f'(0)=1$,
$f'(\exp(it))\ne 0$; the Taylor coefficients $c_1, c_2, c_3, c_4,\ldots
c_k,\ldots$ of a function $f(z)=z+c_1z^2+c_2z^3+c_3z^4+\ldots$ determine a
coordinate system on $S$; necessary and sufficient conditions for the
univalency of a function $f(z)$, which describe the domain $S$ in the linear
space $\Bbb C[[z]]$, are contained in [12-14]. In 1986 A.A.Kirillov [6] (see
also [7]) showed that the class $S$ maybe identified in a natural way with
$\Diff_+(\circle)/\circle$, namely for each element $f$ from $S$ there is
defined the unique function $g$ holomorphic in the exterior of the unit disc,
mapping it onto the exterior of the contour $f(\circle)$ and normalized by the
conditions $g(\infty)=\infty$, $g'(\infty)>0$; let us denote by $\gamma_f$ the
diffeomorphism of the unit circle defined by the formula $\gamma_f=f^{-1}\circ
g$, then the correspondence $f\longmapsto \gamma_f$ defines a bijection of $S$
onto $M(\Vir)$; let us construct the inverse mapping from $M(\Vir)$ onto $S$;
for each diffeomorphism let us consider a manifold $\overline{\Bbb
C}_\gamma=\overline{D}_+\sqcup_\gamma\overline{D}_-$, which is diffeomorphic
and, therefore, boholomorphic to $\overline{\Bbb C}$; let us normalize the
biholomorphic mapping $F:\overline{\Bbb C}_\gamma\mapsto\overline{\Bbb C}$ by
the conditions $F(0)=0$, $F'(0)=1$, $F(\infty)=\infty$ and define $f_\gamma$
from $S$ corresponding to the diffeomorphism $\gamma$ as $\left.
F\right|_{D_+}$. The action of $\CVect(\circle)$ on $\Diff_+(\circle)/\circle$
in the coordinate system $\{c_k\}$ has the form
$$\align
\Cal
L_vf(z)&=-if^2(z)\oint[\frac{wf'(w)}{f(w)}]^2\frac{v(w)}{f(w)-f(z)}\frac{dw}{w};
\\
L_p&=\pd{c_p}+\sum_{k\ge 1}(k+1)c_k\pd{c_{k+p}}\quad (p>0),\\
L_0&=\sum_{k\ge 1}kc_k\pd{c_k},\\
L_{-1}&=\sum_{k\ge 1}((k+2)c_{k+1}-2c_1c_k)\pd{c_k},\\
L_{-2}&=\sum_{k\ge 1}((k+3)c_{k+2}-(4c_2-c^2_1)c_k-b_k(c_1,\ldots
c_{k+2}))\pd{c_k},\\
L_{-n}&=\frac 1{(n-2)!}\ad^{n-2}L_{-1}\cdot L_{-2}\quad (n>2),
\endalign
$$
where $b_k$ are the Laurent coefficients of the function $1/(wf(w))$; the
connection of the described formulas with the classical variational formulas
of the theory of univalent functions is described in [8,16-18]. The symplectic
structures $\omega_{a,b}$ coupled with the complex structure on $M(\Vir)$ form
the two--parameter family of K\"ahler matrics $w_{a,b}$; the more detailed
information on the K\"ahler geometry on the flag manifold for the
Virasoro--Bott group maybe received from the original papers [6-8,17,19] or
the vocabulary [18].

In 1968 I.M.Gelfand and D.B.Fuchs discovered a non--trivial 3--cocycle of Lie
algebra $\Vect(\circle)$, it maybe written as $c(u,v,w)=\det(B(t_0))$, where
$B(t)=\left[\matrix u'(t) & v'(t) & w'(t) \\ u''(t) & v''(t) & w''(t) \\
u'''(t) & v'''(t) & w'''(t) \endmatrix\right]$, or as $c(u,v,w)=\int
\det(B(t))\, dt$ [20]. This cocycle defines a nonassociative deformation of
the Virasoro--Bott group, a loop [21-25], which will be called Gelfand--Fuchs
loop and will be denoted by $\Gf$. the construction of such loop is deeply
related to Finsler geometry [26,27] on $M(\Vir)$. Namely, the Gelfand--Fuchs
3--cocycle defines a closed 3--form $A_{\alpha\beta\gamma}$ on the flag
manifold $M(\Vir)$ for the Virasoro--Bott group. One should consider the
Finsler form
$\Omega{\alpha\beta}=w_{\alpha\beta}+A_{\alpha\beta\gamma}\xi^{\gamma}$, where
$A=dw$, $\xi^\gamma\in T(M(\Vir))$. Then there exists a Finsler connection in
a line holomorphic bundle over $M(\Vir)$ whose curvature is equal to
$\Omega_{\alpha\beta}$. The elements of the group $\Diff_+(\circle)$ are
lifted with respect to this finsler connection to the elements of its
one--dimensional nonassociative central extension which is just the
Gelfand--Fuchs loop $\Gf$.

\subhead 1.2. Non--Euclidean geometry of mirrors on the flag manifold
$M(\Vir)$ for the Virasoro--Bott group, Krichever mapping, skeleton of calss
$S$ of univalent functions, the space of angles $\Ang(\Vir)$ and the integral
geometry on the infinite dimensional non--Euclidean oriented symplectic
grassmannian $\Lambda_+(\Vir)$
\endsubhead

Points and Lagrange submanifolds maybe considered as basic elements of
symplectic geometry. The space of all Lagrange submanifolds of symplectic
geometry. the space of all Lagrange submanifolds is however as a rule
ill--visible. The K\"ahler geometry on the flag manifold $M(\Vir)$ for the
Virasoro--bott group permits to single out a reasonable object from the set of
all Lagrange submanifolds. By a K\"ahler subsymmetry we mean an involutory
antiautomorphism of $M(\Vir)$. The set of all fixed points of a subsymmetry
(mirror) is a completely geodesic Lagrange submanifold. Points and mirrors are
the basic elements of the geometry being described [18,28,29]. The set of all
mirrors forms a symmetric space, the infinite dimensional non--Euclidean
symplectic grassmannian $\Lambda(\Vir)$, independent of a choice of the
non--Einshteinian K\"ahler $\Diff_+(\circle)$--invariant metric on $M(\Vir)$;
subsymmetries are conjugate to the standard subsymmetry $s_-(z)=\bar z$. Let
us consider the probabilistic relization of $M(\Vir)$ and the set $A(M(\Vir))$
of measures of the form $\delta_a(t)dt$; the set $A(M(\Vir))$, which is called
the absolute of $M(\Vir)$, is isomorphic to $\circle$. Let us introduce a
relation of parallelism on the infinite dimensional non--Euclidean symplectic
grassmannian $\Lambda(\Vir)$: two mirrors will be said to be parallel if they
go through the same point of the absolute $A(M(\Vir))$; the following analogue
of the Lobachevskii axiom holds: exactly one mirror passes through any point
of $M(\Vir)$ and any point of $A(M(\Vir))$. let us also introduce the infinite
dimensional non--Euclidean oriented symplectic grassmannian $\Lambda_+(\Vir)$:
elements of $\Lambda_+(\Vir)$ are the oriented mirrors $V_a$, pairs $(V,a)$,
where $V$ is a mirror form $\Lambda(\Vir)$, and $a$ is a point of the
absolute, which belongs to $V$. More detailed information on the infinite
dimensional non--Euclidean geometry of mirrors maybe received from the papers
[18,28,29].

For the following purposes we shall consider an equivariant mapping of the
flag manifold for the Virasoro--Bott group into the infinite dimensional
classical domain of the third type [16-18,4]; let $H$ be the completion of the
space of the smooth real--valued 1--forms $u(\exp(it))dt$ on a circle such
that $\int u(\exp(it))dt=0$ by the norm $||u||^2=\sum |u_n|^2/n$; let $H^{\Bbb
C}$ be its complexification, $H^{\Bbb C}_{\pm}$ be the transversal spaces,
consisting of 1--forms $u(\exp(it))dt$, which maybe holomorphically extended
to the discs $D_{\pm}$; $H^{\Bbb C}\simeq\Cal O(\circle)/\Const$, namely
$f(z)\in\Cal O(\circle)\mapsto df(z)\in H^{\Bbb C}$, $H^{\Bbb
C}_{\pm}\simeq\Cal O(D_{\pm})/\Const$; there are the symplectic and the
pseudohermitean structures defined on $H^{\Bbb C}$: $(f(z),g(z))=\int
f(z)\,dg(z)$, $<f(z),g(z)>=\int f(z)\,\overline{dg(z)}$ ($f,g\in\Cal
O(\circle)$); let $\Sp(H^{\Bbb C},\Bbb C)$ and $U(H^{\Bbb C}_+,H^{\Bbb C}_-)$
be the groups of invariance of these structures, $\Sp(H,\Bbb R)=\Sp(H^{\Bbb
C},\Bbb C) U(H^{\Bbb C}_+,H^{\Bbb C}_-)$. Let us consider the grassmannian
$\Gr(H^{\Bbb C})$ -- the set of all complex Lagrange subspaces in $H^{\Bbb
C}$, $\Gr(H^{\Bbb C})$ is an infinite dimensional homogeneous space with the
group of transformations $\Sp(H^{\Bbb C},\Bbb C)$. Let us consider the action
of $\Sp(H,\Bbb R)$ on $\Gr(H^{\Bbb C})$; the orbit of the point $H^{\Bbb C}_-$
is an open subspace $R$ in $\Gr(H^{\Bbb C})$ isomorphic to $\Sp(H,\Bbb R)/U$,
where $U=\{A\oplus\bar A, A\in U(H^{\Bbb C}_+), \bar A\in U(H^{\Bbb C}_-)\}$,
the space $R$ is an infinite dimensional classical homogeneous domain of the
third type (i.e. an infinite dimensional analogue of finite dimensional
classical domains of this type [30]), $R$ is mapped in the linear space
$\Hom(H^{\Bbb C}_+,H^{\Bbb C}_-)$ so that the elements of $R$ are represented
by symmetric matrices $Z$ such that $E-Z\bar Z>0$. The detailed information
on infinite dimensional grassmannians maybe received from the papers [31-33];
the construction of the mapping of $M(\Vir)$ into $R$ was described in the
papers [16-18,4], namely, the representation of $\Diff_+(\circle)$ in $H$
defines a monomorphism $\Diff_+(\circle)\mapsto\Sp(H,\Bbb R)$, hence
$\Diff_+(\circle)$ acts on $R$; the orbit of the initial point under this
action coincides with $\Diff_+(\circle)/\PSl(2,\Bbb R)$, therefore we have an
equivariant mapping of $M(\Vir)$ into $R$, its explicit form can be found in
[16-18]. The matrix $Z_f$ from $R$ which corresponds to the function $f\in S$
is called the Grunsky matrix [34] (the definitions of the grunsky matrix maybe
found also in [14,16]), the mapping $f\mapsto Z_f$ is the partial case of the
Krichever mapping [32,35].

The skeleton of the domain $R$ consists of all symmetric unitary matrices $Z$,
therefore, the skeleton of $S=\Diff_+(\circle)/\circle$ consists of univalent
functions which Grunsky matrices are unitary. accordingly to the Milin theorem
[36] the skeleton of the space $S$ consists of all univalent functions $f$
such that $\mes(\Bbb C\backslash(D^{\circ}_+))=0$, but the action of the
group $\Diff_+(\circle)$ on the skeleton is not transitive; the structure of
the skeleton of class $S$ is described in the paper [29]. Following it let us
consider the $\Bbb R$--analytic manifold $E$ whose elements are cuts $K$ of
the complex plane $\Bbb C$ with one end at infinity, such that the conformal
radius of $\Bbb C\backslash K$ is equal to one. Let us consider the mapping
$E\mapsto\Lambda_+(\Vir)$, defined as $f(z)\mapsto (s,a)$, where
$f(D^\circ_+)\sqcup K=\Bbb C$, $f(0)=0$, $f'(0)=1$, $f(a)=\infty$,
$f(s(z))=f(z)$; it was shown in the paper [29] that this mapping is an
isomorphism. let us mention that the action of $\Diff_+(\circle)$ on $S$ maybe
analytically extended on $E$, so that $E$ and $\Lambda_+(\Vir)$ are isomorphic
as homogeneous spaces. It should be mentioned that the infinite dimensional
non--Euclidean oriented symplectic grassmannian $\Lambda_+(\Vir)$ is a
symmetric space: $(s_1,a_1)\circ(s_2,a_2)=(s_1s_2s_1,s_1(a_2))$ with the group
of transvections $\Diff_+(\circle)$ and the isotropy group
$G_0=\{g\in\Diff_+(\circle):g(\bar z)=\overline{g(z)}, g(1)=1\}$, the tangent
space $V$ to $\Lambda_+(\Vir)$ at the point $(s_-,1)$ maybe identified with
the space of odd vector fields on $\circle$. The lines in $V$ invariant with
respect to $G_0$, which are determined by the generalized vectors
$\delta_{\pm}(t)d/dt$, correspond to the nonholonomic generalized invariant
fields $\xi_{\pm}$ on $\Lambda_+(\Vir)$; let $\Cal O(E)$ and $\Cal
O(\Lambda_+(\Vir))$ be the structural rings of $E$ and $\Lambda_+(\Vir)$;
$\Cal O(\Lambda_+(\Vir)/\xi_-)=\{f\in\Cal O(\Lambda_+(\Vir)):\xi_-f=0\}$; the
isomorphism of the ringed spaces holds: $(E,\Cal
O(E))\simeq(\Lambda_+(\Vir),\Cal O(\Lambda_+(\Vir)/\xi_-))$ [29].

\define\real{\operatorname{Re}}
\define\cl{\operatorname{cl}}
\define\hol{\operatorname{hol}}
\define\nh{\operatorname{nh}}
Let us consider the mapping $M(\Vir)\mapsto \Gamma_{\cl}(\Lambda_+(\Vir))$,
where $\Gamma_{\cl}(\Lambda_+(\Vir))$ is the space of all closed geodesics on
$\Lambda_+(\Vir)$, namely, we shall assign to a point $x$ the set of all
oriented mirrors which pass through it; this mapping is an isomorphism [29].
Under the identification with $\Gamma_{\cl}(\Lambda_+(\Vir))$ the symplectic
structure on $M(\Vir)$ has the form $\omega_x(X,Y)=\int_{\gamma_x}(AX,Y)\,ds$,
where $s$ is a natural parameter on $\gamma_x$, $X$ and $Y$ are the Jacobi
fields orthogonal to the field $\dot s$ for the unique (up to a multiplication
by a real number) invariant degenerate pseudoriemannian metric on
$\Lambda_+(\Vir)$, $A=a\nabla_s+b\nabla_s^3$, where $\nabla_s$ is the
covariant derivative along $\dot s$. Let $\Cal O^\circ(S)$ be the class of all
holomorphic functionals on $S$ which admit an analytic extension to $E$ then
$\real \Cal O^\circ(S)\simeq\Cal O(\Lambda_+(\Vir)/\xi_-)$, namely [29]
$F(f)=\int_{f_s\!\in\!\gamma_x}F(f_s)\,ds$ (**). Let us denote by
$\Lambda_+^{\nh}(\Vir)$ the nonholonomic manifold (cf.[37]) the "spectrum" of
$\Cal O(\Lambda_+(\Vir)/\xi_-)$, let
$\Cal O_{\hol}(\Lambda^{\nh}_+(\Vir))$ be the inverse image of $\Cal
O^\circ(M(\Vir))$ in $\Cal O^{\Bbb C}(\Lambda^{\nh}_+(\Vir))$ under the
mapping (**).

\define\irr{\operatorname{irr}}
An angle $\hat\alpha$ on the flag space $M(\Vir)$ is a triple
$\hat\alpha=(x,U,V)$, $x\!\in\! M(\Vir)$, $U,V\!\in\!\Lambda_+(\Vir)$,
$x\!\in\! U\cap V$, the point $x$ is called the vertex of an angle and the
mirrors $U$ and $V$ are called the sides of an angle $\hat\angle$; for two
angles with a common vertex and a common side there is defined a sum
$\hat\alpha+\hat\beta=(x,U,W)$, where $\hat\alpha=(x,U,V)$,
$\hat\beta=(x,V,W)$; two angles are called congruent if they maybe mapped one
into another by some element of the group $\Diff_+(\circle)$, which acts on
$M(\Vir)$. This action conserves an additive invariant of angles --- their
value, the real number defined up to $2\pi k$, $k\!\in\!\Bbb Z$; the angles
with the vertex $x$ form a compact riemannian manifold --- two--dimensional
torus $\Bbb T^2$; an angle is called rational if and only if its value is a
rational number and irrational otherwise; an angle $\hat\alpha=(x,U,V)$ is
irrational if and only if $U\cap V=\{x\}$. The space of angles $\Ang(\Vir)$ is
a bibundle:
$M(\Vir)\longleftarrow\Ang(\Vir)\longrightarrow\Lambda_+(\Vir)\times\Lambda_+(\Vir)$
($\pi_1:\Ang(\Vir)\mapsto M(\Vir)$,
$\pi_2:\Ang(\Vir)\mapsto\Lambda_+(\Vir)\times\Lambda_+(\Vir)$); thereis
defined a mapping $\pi_1\circ\pi_2^{-1}$ on $\pi_2(\Ang^{\irr}(\Vir)$, where
$\Ang^{\irr}(\Vir)$ is the space of irrational angles; the image of this
mapping coincides with $M(\Vir)$. let us consider a mapping from $\Cal
O_{\hol}(\Lambda^{\nh}_+(\Vir)\times\Lambda^{\nh}_+(\Vir))$ to $\Cal
O(M(\Vir))$ defined as $\hat F(x)=\int_{\hat\alpha:\pi_1(\hat\alpha)=x}
F(\pi_2(\hat\alpha))\,d\mu_{\Bbb T^2}$, where $d\mu_{\Bbb T^2}$ is the
canonical measure on torus $\Bbb T^2$.

\proclaim{\rm THEOREM 1} The mapping $F\mapsto\hat F$ is injective and has a
dense
image in $\Cal O(M(\Vir))$ after the restriction on the solutions of the
system of equations $\square_p F=0$, $(X^{(1)}-X^{(2)})F=0$
($\square_p=\sum_{i+j=p}(i-j)(L_i^{(1)}L_j^{(2)}-L_j^{(1)}L_i^{(2)})$, $X$ is
the vector field that generate a geodesics passing through points 1 and 2).
\endproclaim

The statement of the theorem is an evident consequence of the integral
formulas (**).

\subhead 1.3. Verma modules over the Virasoro algebra (Feigin--Fuchs theory
and orbit method) \endsubhead

The algebras $\CVect(\circle0$ and $\Cvir$ are $\Bbb Z$--graded:
$\deg(e_k)=k$, for them there are defined Verma modules, which ewre
investigated by several authors [38--40]. namely, let $\Cvir^+=\span(e_k,k\ge
0)$, $\chi_{h,c}$ be the character of $\Cvir$, defined by the conditions
$\chi_{h,c}(e_k)=0$ if $k>0$, $\chi_{h,c}(e_0)=h$, $\chi_{h,c}(c)=c$; the Verma
module $V_{h,c}$ is the $\Cvir$--module induced by the character $\chi_{h,c}$
of the subalgebra $\Cvir^+$; otherwords, $V_{h,c}=\Cal U(\Cvir)\otimes_{\Cal
U(\Cvir^+)}V_{\chi_{h,c}}$, where $V_{\chi_{h,c}}$ is the $\Cvir$--module
defined by the character $\chi_{h,c}$, $\Cal U(\Cvir)$ and $\Cal U(\Cvir^+)$
are the universal envelopping algebras of Lie algebras $\Cvir$ and $\Cvir^+$.
the Verma module $V_{h,c}$ is a graded $\Cvir$--module; if $h$ and $c$ are
real (that will be supposed below) there is defined the unique up to a
multiple invariant hermitean form on $V_{h,c}$. the Verma module $V_{h,c}$ is
unitarizable if and only if the hermitean form is positive definite; let us
denote by $D_n(h,c)$ the determinant of this form in $n$-th homogeneous
component of $V_{h,c}$ in the basis $L_{-1}^{k_1}\ldots L_{-j}^{k_j}v$,
$k_j\ge 0$ ($L_0 v=hv$, $L_m v=0$), then as it aws shown by V.G.kac,
B.L.Feigin and D.B.Fuchs $D_n(h,c)=A\prod_{0<\alpha\le\beta, \alpha\beta\le
n}\Phi^{p(n-\alpha\beta)}_{\alpha,\beta}$, where
$\Phi_{\alpha,\alpha}(h,c)=h+\frac{c-13}{24}(\alpha^2-1_)$,
$\Phi_{\alpha,\beta}(h,c)=(h+\frac{c-13}{24}(\beta^2-1)+
\frac{\alpha\beta-1}2)(h+\frac{c-13}{24}(\alpha^2-1)+\frac{\alpha\beta-1}2)+
\frac{\alpha^2-\beta^2}{16}$. If for any $\alpha$, $\beta$
$\Phi_{\alpha,\beta}(h,c)\ne 0$ then the module $V_{h,c}$ is irreducible and
is not contained in any other Verma module; if there exist exactly one pair
$\alpha,\beta$ such that $\Phi_{\alpha,\beta}(h,c)=0$ then three
possibilities maybe realized: 1) $\alpha\beta<0$, then $V_{h,c}$ maybe
imbedded into the Verma module $V_{h+\alpha\beta,c}$, 2) $\alpha\beta>0$, then
$V_{h,c}$ contains a submodule $V_{h+\alpha\beta,c}$, 3) $\alpha=0$ or
$\beta=0$, then $V_{h,c}$ is irreducible and is not a submodule of another
Verma module; if there exist two pairs $(\alpha_1,\beta_1)$ and
$(\alpha_2,\beta_2)$ such that $\Phi_{\alpha_i,\beta_i}(h,c)=0$, then there
exist an infinite number of pairs $(\alpha,\beta)$, which possess such
property --- this situation is realized if
$$\align
c_{1,2}&=1-\frac{6((\alpha_1\pm\alpha_2)-(\beta_1\pm\beta_2))^2}
{(\alpha_1\pm\alpha_2)(\beta_1\pm\beta_2)}\tag 1A\\
h_{1,2}&=\frac{(\alpha_2\beta_1-\alpha-1\beta_2)^2-((\alpha_1\pm\alpha_2)-
(\beta_1\pm\beta_2))^2}{4(\alpha_1\pm\alpha-2)(\beta_1\pm\beta_2)}\tag 1B
\endalign
$$
In this case the structure of the Verma modules is described by the
Feigin--Fuchs theory.

The Verma module $V_{h,c}$ is unitarizable if $h>0$, $c>1$; the Verma module
$V_{h,c}$ contains an unitarizable quotient if (a) $h>0$, $c>1$; (b)
$c=1-\frac6{p(p+1)}$, $h=\frac{(\alpha p-\beta(p+1))^2-1}{4p(p+1)}$,
$\alpha,\beta,p\!\in\!\Bbb Z$; $p\ge 2$; $1\le\alpha\le p$, $1\le\beta\le
p+1$.

Let us describe a geometric way of the construction of the Verma modules over
the Virasoro algebra, based on the orbit method; it is described by the
following facts [16--18]: (1) To each $\Diff_+(\circle)$--invariant K\"ahler
metric $w_{h,c}$ on the space $M(\Vir)$ one should correspond the linear
holomorphic bundle $E_{h,c}$ over $M(\Vir)$ with the following properties: (a)
$E_{h,c}$ is the hermitean bundle with the metric $\exp(-U_{h,c})d\lambda
d\bar\lambda$, where $\lambda$ is a coordinate in a fiber,
$K_{h,c}=\exp(U_{h,c})$ is the Bergman kernfunction, the exponential of the
K\"ahler potential $U_{h,c}=h\log|g'(\infty)|-c\log\det(E-Z_f\bar Z_f)$ of the
metric $w_{h,c}$, (b) algebra $\Cvir$ holomorphically acts in the prescribed
bundle by covariant derivatives with respect to the hermitean connection with
the curvature form being equal to $2\pi\omega_{h,c}$; (2) let $\Cal
O(E_{h,c})$ be the space of all polynomial (in the trivialization of the paper
[17]) germs of sections of the bundle $E_{h,c}$, $\Cal O(E_{h,c})$ is the
graded $\Cvir$--module, the action of the Lie algebra $\Cvir$ in which is
defined by the following formulas [17,18]
$$\allowdisplaybreaks\align
L_p&=\pd{c_p}+\sum_{k\ge 1}(k+1)c_k\pd{c_{k+p}}\quad (p>0),\\
L_0&=\sum_{k\ge 1}kc_k\pd{c_k}+h,\\
L_{-1}&=\sum_{k\ge 1}((k+2)c_{k+1}-2c_1c_k)\pd{c_k}+2hc_1,\\
L_{-2}&=\sum_{k\ge 1}((k+3)c_{k+2}-(4c_2-c_1^2)c_k-b_k(c_1,\ldots
c_{k+2}))\pd{c_k}+h(4c_2-c_1^2)+\frac{c}2(c_2-c_1^2),\\
L_{-n}&=\frac1{(n-2)!}\ad^{n-2}L_{-1}\cdot L_{-2}\quad (n>2);
\endalign
$$
let us fix the basis $e^{a_1\ldots a_n}=c_1^{a_1}\ldots c_n^{a_n}$ in $\Cal
O(E^*_{h,c})$ and let $O^*(E^*_{h,c})$ be the space of all linear functionals
on $\Cal O(E^*_{h,c})$, which obey the next property: if $p(x)>0$, then
$\deg(x)\le N_p$; the space $\Cal O^*(E^*_{h,c})$ is called the Fock space of
the apir $(M(\Vir),E_{h,c})$ [41,p.117] (the Fock spaces plays an essential
role in the method of geometric quantization, the standard Fock space is a
partial case of the spaces introduced in [41]), the Verma module $V_{h,c}$ is
realized in the Fock space $F(E_{h,c})$ of the pair $(M(\Vir),E_{h,c})$; let
us fix the basis $e_{a_1\ldots a_n}=:c_1^{a_1}\ldots c_n^{a_n}$ in
$F(E_{h,c})$ such that $<e_{a_1\ldots a_n},e^{b_1\ldots b_m}>=a_1!\ldots
a_n!\delta^m_n\delta^{b_1}_{a_1}\ldots\delta^{b_n}_{a_n}$; the action of the
Virasoro algebra in such basis is defined by the next formulas [17,18]
$$\allowdisplaybreaks\align
L_{-p}=&c_p+\sum_{k\ge 1}(k+1)c_{k+p}\pd{c_k}\quad (p>0),\\
L_0=&\sum_{k\ge 1}kc_k\pd{c_k}+h,\\
L_1=&\sum_{k\ge 1}c_k((k+2)\pd{c_{k+1}}-2\pd{c_1}\pd{c_k})+2h\pd{c_1},\\
L_2=&\sum_{k\ge
1}c_k((k+3)\pd{c_{k+2}}-(4\pd{c_2}-(\pd{c_1})^2)\pd{c_k}-
b_k(\pd{c_1},\ldots\pd{c_{k+2}}))+\\&h(4\pd{c_2}-(\pd{c_1})^2)+\frac{c}2(\pd{c_2}-
(\pd{c_1})^2),\\
L_n=&\frac{(-1)^n}{(n-2)!}\ad^{n-2}L_1\cdot L_2\quad (n>2).
\endalign
$$

\subhead 1.4. The fundamental affine space $A(\Vir)$ and the $\Pi$--space
$\Pi(\Vir)$ for the Virasoro--Bott group, the model and the skladen of the
Verma modules over the virasoro algebra. The space $\Sigma(\Vir)$ of punctured
oriented mirrors and the model of the Verma modules over the Virasoro algebra.
the space $\Sigma^\#(\Vir)$, the skladen of the Verma modules over the
Virasoro algebra and the Alekssev--Shatashvili construction. The universal
deformation of a complex disc, the Manin--kontsevich--Beilinson--Schechtman
construction and the model of the Verma modules over the Virasoro algebra
\endsubhead

It is well--known from the theory of representations of compact groups that
the model of the representations of a compact Lie group $U$ is realised in the
space of functions on the fundamental affine space for its complexification
$G^{\Bbb C}$; the model of the representations of the group $U$ is a
representation which maybe expanded in the direct sum of the irreducible ones
so that each such representation appear in the sum exactly one time; the
fundamental affine space is a homogeneous space $G^{\Bbb C}/N$ of the complex
reductive group $G^{\Bbb C}$ with the stationary subgroup being isomorphic to
the maximal nilpotent subgroup $N$. An analogous construction holds for the
real noncompact simple Lie groups of Hermitean type (there is an inaccuracy in
the paper [41]: the supposition of caspidality should be changed on the
condition formulated above), namely, let $G$ be a real noncompact simple Lie
group of Hermitean type, $G^{\Bbb C}$ be its complexification, $\Mantle(G)$ be
the semigroup lying in $G^{\Bbb C}$ (so--called partial complexification of
the group $G$ [42,43]); the quotient $\Mantle(G)/\Mantle(G)\cap N$ is called
the fundamental affine space for the Lie group $G$; the fundamental affine
space of the universal covering $\tilde G$ of the group $G$ is the universal
covering of its fundamental affine space. as it was shown in the paper [41] in
the Fock space of the pair (the fundamental affine space of the group $G$,
trivial holomorphic bundle over it) there is realised a direct integral of the
Verma modules over the Lie algebra $\frak g^{\Bbb C}$ (the model of the Verma
modules over it). In the case of the Virasoro--Bott group $\Vir$ the
corresponding complex group $\CVir$ does not exist, but the semigroup
$\Mantle(\Vir)$, the Neretin semigroup $\Ner$ exists, so the construction of
the fundamental affine space maybe extended on the infinite dimensional case;
the fundamental affine space of the group $\Vir$ consists of the pairs
$(f,t)$, where $f$ is an univalent function from $S$, and $t$ is a non--zero
complex number, $|t|>1$ [41]; the Virasoro algebra generators have the form
[41]
$$\allowdisplaybreaks\align
L_p=&\pd{c_p}+\sum_{k\ge 1}(k+1)c_k\pd{c_{k+p}}\quad (p>0),\\
L_0=&\sum_{k\ge 1}kc_k\pd{c_k}+t\pd{t},\\
L_{-1}=&\sum_{k\ge 1}((k+2)c_{k+1}-2c_1c_k)\pd{c_k}+2c_1t\pd{t},\\
L_{-2}=&\sum_{k\ge 1}((k+3)c_{k+2}-(4c_2-c_1^2)c_k-b_k(c_1,\ldots
c_{k+2}))\pd{c_k}+(4c_2-c_1^2)t\pd{t},\\
L_{-n}=&\frac1{(n-2)!}\ad^{n-2}L_{-1}\cdot L_{-2}\quad (n>2).
\endalign
$$
Let us consider the bundle $E(c)$ over the fundamental affine space, the
inverse image of the bundle $E_{0,c}$ under the projection onto the flag
manifold; the Virasoro algebra generators in the Fock space of the bundle
$E(c)$ have the form [41]
$$\allowdisplaybreaks\align
L_{-p}=&c_p+\sum_{k\ge 1}(k+1)c_{k+p}\pd{c_k}\quad (p>0),\\
L_0=&\sum_{k\ge 1}kc_k\pd{c_k}-t\pd{t},\\
L_1=&\sum_{k\ge 1}c_k((k+2)\pd{c_{k+1}}-2\pd{c_1}\pd{c_k})-2t\pd{t}\pd{c_1},\\
L_2=&\sum_{k\ge 1}c_k((k+3)\pd{c_{k+2}}-(4\pd{c_2}-(\pd{c_1})^2)\pd{c_k}-
b_k(\pd{c_1},\ldots\pd{c_{k+2}}))-\\
&t\pd{t}(4\pd{c_2}-(\pd{c_1})^2)+\frac{c}2(\pd{c_2}-(\pd{c_1})^2),\\
L_n=&\frac{(-1)^n}{(n-2)!}\ad^{n-2}L_1\cdot L_2\quad (n>2).
\endalign
$$
The model of the Verma modules over the Virasoro algebra is realised in the
Fock space of the bundle $E(c)$ over the fundamental affine space for the
group $\widetilde {\Vir}$.

\define\PSltwo{\operatorname{PSl}(2,\Bbb R)}
\define\cPSltwo{\widetilde{\operatorname{PSl}}(2,\Bbb R)}
\define\can{\operatorname{can}}
As it was mentioned in the paper [44] the skladen of the Verma modules over
the Lie algebra $\sLtwo$ is realised in the Fock space over the $\Pi$--space
$\Pi(\cPSltwo)$ for the universal covering $\cPSltwo$ of the projective group
$\PSltwo$; the skladen of the Verma modules over the simple complex Lie
algebra $\frak g^{\Bbb C}$ is a direct integral $\int_W V_{\pi(x)}\,dx$ of the
Verma modules over this algebra of the weight $\pi(x)$, where $W=\frak
h\oplus\frak h^*$, $\pi:W\mapsto\frak h^*$, $\frak h$ is the Cartan subalgebra
in $\frak g^{\Bbb C}$, $\frak h^*$ is the weight space; the $\Pi$--space of
the real simple Lie group $G$ of Hermitean type maybe defined by the
noncommutative diagram

$$\align
&\quad\Bbb C^r\quad\,\,\simeq\quad\,\,\Bbb C^r\\
&\quad\downarrow\qquad\qquad\,\,\downarrow\\
&T^*([D^*_+]^r)\mapsto\Pi(G)\mapsto M(G)\\
&\quad\downarrow\qquad\qquad\,\,\downarrow\qquad\qquad\parallel\\
&\quad[D^*_+]^r\,\,\mapsto\,\, A(G)\mapsto M(G)
\endalign$$
the $\Pi$--space of the universal covering $\tilde G$ of the Lie group $G$ is
the universal covering of the $\Pi$--space $\Pi(G)$. The $\Pi$--space of the
group $\Vir$ consists of triples $(f,t,s)$, where $f$ is an univalent function
from $S$, $t$ is a non--zero complex number, $|t|<1$, $s$ is an arbitrary
complex number (or an element of the universal covering $\tilde{\Bbb C}^*$ of
the complex plane without zero; the Virasoro algebra generators have the form
$$\allowdisplaybreaks\align
L_p=&\pd{c_p}+\sum_{k\ge 1}(k+1)c_k\pd{c_{k+p}}\quad (p>0),\\
L_0=&\sum_{k\ge 1}kc_k\pd{c_k}+t\pd{t}+s\pd{s},\\
L_{-1}=&\sum_{k\ge 1}((k+2)c_{k+1}-2c_1c_k)\pd{c_k}+2c_1t\pd{t}+2c_1s\pd{s},\\
L_{-2}=&\sum_{k\ge 1}((k+3)c_{k+2}-(4c_2-c_1^2)c_k-b_k(c_1,\ldots
c_{k+2}))\pd{c_k}+(4c_2-c_1^2)t\pd{t}+(4c_2-c_1^2)s\pd{s},\\
L_{-n}=&\frac1{(n-2)!}\ad^{n-2}L_{-1}\cdot L_{-2}\quad (n>2).
\endalign
$$
Let us consider the bundle $E(c)$ over the $\Pi$--space $\Pi(\Vir)$, the
inverse image of the bundle $E(c)$ over the fundamental affine space $A(\Vir)$
under the projection of $\Pi(\Vir)$ on $A(\Vir)$; the Virasoro algebra
generators in the Fock space of the bundle $E(c)$ have the form
$$\allowdisplaybreaks\align
L_{-p}=&c_p+\sum_{k\ge 1}(k+1)c_{k+p}\pd{c_k}\quad (p>0),\\
L_0=&\sum_{k\ge 1}kc_k\pd{c_k}-t\pd{t}-s\pd{s},\\
L_1=&\sum_{k\ge 1}c_k((k+2)\pd{c_{k+1}}-2\pd{c_1}\pd{c_k})-2t\pd{t}\pd{c_1}
-2s\pd{s}\pd{c_1},\\
L_2=&\sum_{k\ge 1}c_k((k+3)\pd{c_{k+2}}-(4\pd{c_2}-(\pd{c_1})^2)\pd{c_k}-
b_k(\pd{c_1},\ldots\pd{c_{k+2}}))-\\
&t\pd{t}(4\pd{c_2}-(\pd{c_1})^2)-s\pd{s}(4\pd{c_2}-(\pd{c_1})^2)+
\frac{c}2(\pd{c_2}-(\pd{c_1})^2),\\
L_n=&\frac{(-1)^n}{(n-2)!}\ad^{n-2}L_1\cdot L_2\quad (n>2).
\endalign
$$

\proclaim{\rm THEOREM 2} The skladen of the Verma modules over the Virasoro
algebra $\Cvir$ is realised in the Fock space of the pair
$(\Pi(\widetilde{\Vir});E(c))$.
\endproclaim

Namely, the highest weights in this Fock space have the form
$t^h\varphi(t/s)$.

A punctured oriented mirror $V^x_a$ on the flag manifold $M(Vir)$ is a pair
$(V_a,x)$, where $V_a$ is an oriented mirror and $x$ is a point of $M(\Vir)$
which belongs to $V_a$; the space $\Sigma(\Vir)$ of the punctured oriented
mirrors is a bobundle
$M(\Vir)\longleftarrow\Sigma(\Vir)\longrightarrow\Lambda_+(\Vir)$,
where $\Sigma(\Vir)\mapsto M(\Vir)$ is the subsymmetry bundle [28,18] and
$\Sigma(\Vir)\mapsto\Lambda_+(\Vir)$ is the tautological bundle. On the space
$\Sigma(\Vir)$ of the punctured oriented mirrors there is defined a structure
of a contact CR--manifold [45] by the canonical connection of the
prequantisation [28,18] in the subsymmetry bundle, the contact structure is
defined by the connection form $\theta^{\can}$ and the structure of
CR--manifold on the space $\Sigma(\Vir)$ of the punctured oriented mirrors is
defined by the complex structure on horisontal subspaces of the subsymmetry
bundle lifted from the base of the bundle, the flag manifold $M(\Vir)$; the
Levy form of the CR--manifold $\Sigma(\Vir)$ coincides with the K\"ahler form
on the base. Let us denote by $\Sigma(\widetilde{\Vir})$ the universal
covering of the space $\Sigma(\widetilde{\Vir})$. Let us denote by E(c) the
CR--analytic bundle over the space $\Sigma(\widetilde{\Vir})$ of the punctured
oriented mirrors, lifted from the bundle $E_{o,c}$ on the base.

\proclaim{\rm THEOREM 3} The model of the Verma modules over the Virasoro
algebra
is realised in the Fock space of the pair $(\Sigma(\widetilde{\Vir}),E(c))$.
\endproclaim

The statement of the theorem is the consequence of the result obtained in the
third paragraph of the paper [41].

Let us consider the space $\Sigma^\#(\Vir)$ defined by the commutative diagram

$$\align
&\,\,\quad\Bbb R\quad\,\,\simeq\quad\,\,\Bbb R\\
&\quad\downarrow\qquad\qquad\,\,\downarrow\\
&T^*(\operatorname{\Bbb RP}^1)\mapsto\Sigma^\#(\Vir)\mapsto M(\Vir)\\
&\quad\downarrow\qquad\qquad\,\,\downarrow\qquad\qquad\,\,\parallel\\
&\quad\operatorname{\Bbb RP}^1\,\,\mapsto\,\, A(\Vir)\,\,\mapsto\, M(\Vir)
\endalign$$

On $\Sigma^\#(\Vir)$ there is defined a structure of the CR--manifold; let us
denote by $\Sigma^\#(\widetilde{\Vir})$ the universal covering of the space
$\Sigma^\#(\Vir)$, and by $E(c)$ the CR--analytic bundle over
$\Sigma^\#(\widetilde{\Vir})$ lifted from $\Sigma(\widetilde{\Vir})$.

\proclaim{\rm THEOREM 4}
The skladen of the Verma modules over the Virasoro algebra is realised in the
Fock space of the pair $(\Sigma^\#(\widetilde{\Vir}), E(c))$.
\endproclaim

The statement of the theorem follows from the theorem 1.

\define\AS{\operatorname{AS}}
On $\Sigma^\#(\widetilde{\Vir})$ there is defined a structure of symplectic
manifold; the symplectic form on $\Sigma^\#(\widetilde{\Vir})$, the
Alekseev--Shatashvili form $\omega_{\AS}$ maybe obtained by the hamiltonian
reduction of the canonical symplectic structure on $T^*(\cPSltwo)$ [46,47]. In
the papers [46,47] there is considered a procedure of the quantisation of such
symplectic structure, the value of the central charge of the corresponding
bundle $E(c)$ maybe determined from the following characteristics of the
Alekseev--Shatashvili form: thre is defined a fiberwise action of the abelian
group $\Bbb Z^2$ in the bundle $\Bbb
R^2\longleftarrow\Sigma^\#(\widetilde{\Vir})\longrightarrow M(\Vir)$, the
volume $\alpha$ of the fundamental domain in the fiber determines the central
charge $c=1-6(\alpha-\alpha^{-1})^2$; the condition of rationality of such
volume extract the values of central charge from the spectrum (2A) [47]. the
attentive reading of the papers of the mentioned authors shows that the
starting point for the receiving of the Kac spectrum (2) in them is the
skladen more than the model of the verma modules over the Virasoro algebra,
this circumstance explains the appearing of two parameters in Kac spectrum.

Let us consider the representation of elements of the homogeneous space
$\Diff_+(\circle)/\circle$ by univalent functions $f$ from class $S$; the
function $f(1-tf)^{-1}$ belongs to class $S$, if $t^{-1}$ does not belong to
the image of $f$; the set of such points $t$ forms a domain $\Bbb
C\backslash(f(D_+))$ in the complex plane $\Bbb C$. the family of pairs
$(f,t)$ is the space of the universal deformation of a complex disc
[41,p.217],
$S$ is the universal Teichm\"uller space [48-51]; the action of the Virasoro
algebra $\Cvir$ on pairs $(f,t)$ is defined by formulas
$$\allowdisplaybreaks\align
L_p=&\pd{c_p}+\sum_{k\ge 1}(k+1)c_k\pd{c_{k+p}}\quad (p>0),\\
L_0=&\sum_{k\ge 1}kc_k\pd{c_k}+t\pd{t},\\
L_{-1}=&\sum_{k\ge 1}((k+2)c_{k+1}-2c_1c_k)\pd{c_k}+2c_1t\pd{t}+t^2\pd{t},\\
L_{-2}=&\sum_{k\ge 1}((k+3)c_{k+2}-(4c_2-c_1^2)c_k-b_k(c_1,\ldots
c_{k+2}))\pd{c_k}+(4c_2-c_1^2)t\pd{t}+3c_1t^2\pd{t}+t^3\pd{t},\\
L_{-n}=&\frac1{(n-2)!}\ad^{n-2}L_{-1}\cdot L_{-2}\quad (n>2);
\endalign
$$
this action maybe obtained by the Manin--Kontsevich--Beilinson--Schechtman
construction [52-57]; the Virasoro algebra generators in the Fock space of the
bundle $E(c)$ over the universal deformation of a complex disc have the form
$$\allowdisplaybreaks\align
L_{-p}=&c_p+\sum_{k\ge 1}(k+1)c_{k+p}\pd{c_k}\quad (p>0),\\
L_0=&\sum_{k\ge 1}kc_k\pd{c_k}-t\pd{t},\\
L_1=&\sum_{k\ge
1}c_k((k+2)\pd{c_{k+1}}-2\pd{c_1}\pd{c_k})-2t\pd{t}\pd{c_1}+t^2\pd{t},\\
L_2=&\sum_{k\ge 1}c_k((k+3)\pd{c_{k+2}}-(4\pd{c_2}-(\pd{c_1})^2)\pd{c_k}-
b_k(\pd{c_1},\ldots\pd{c_{k+2}}))-\\
&t\pd{t}(4\pd{c_2}-(\pd{c_1})^2)+3t^2\pd{t}\pd{c_1}-t^3\pd{t}+
\frac{c}2(\pd{c_2}-(\pd{c_1})^2),\\
L_n=&\frac{(-1)^n}{(n-2)!}\ad^{n-2}L_1\cdot L_2\quad (n>2);
\endalign
$$
these generators define a structure of the model of the Verma modules over the
Virasoro algebra in the Fock space of the bundle $E(c)$ over the universal
covering of the universal deformation of a complex disc.

\define\DET{\operatorname{DET}}
let us consider the determinant sheaf $\DET_{\lambda}$ over $M(\Vir)$:
$\DET_{\lambda}=R^0p_*(\Omega^{\lambda}(\tilde C/M(\Vir)))$, where $p:C\mapsto
M(\Vir)$ is the projection of the universal deformation of the punctured
complex disc $D^*_+$ onto the base $M(\Vir)$, $\tilde C$ is the universal
covering of $C$ fibred over $M(\Vir)$ with fiber $\tilde D^*_+$,
$\Omega_\lambda$ is the sheaf of holomorphic $\lambda$--differentials on
$\tilde D^*_+$.

\proclaim{\rm THEOREM 5}
The model of the Verma modules over the Virasoro algebra $\Cvir$ with the
central charge $c=2(6\lambda^2-6\lambda+1)$ maybe realised in the space of
sections of the sheaf $\DET_\lambda$.
\endproclaim

The statement of the theorem is a consequence of the fact that the Virasoro
algebra representations obtained by the determinant construction and
non--isomorphic to the Verma modules over the Virasoro algebra have the zero
measure in the space of parameters [39,40], the relation between the central
charge and the parameter $\lambda$ is also described in these papers.

\head 2. Infinite dimensional geometry and quantum field theory of
non--interacting strings: second quantized free string on flat and curved
background, the Bowick--Rajeev formalism of the separation of the internal and
external degrees of freedom of a string
\endhead

The purpose of this chapter is to give a mathematical description of the
second quantization of a closed string based on infinite dimensional geometry.
The configuration space of the quantum field theory is an infinite dimensional
space, its elements are classical fields (functions, distributions,
differential forms, connections) on the support manifold. If these fields are
free of constraints, then the configuration space is flat, and in the presence
of constraints it has a rather complicated structure. if the number of fields
is infinite then the support manifold has an infinite dimension, the
configuration space has a dimension $\infty^2$. This situation is realized in
the case of the closed string--field theory.

\subhead 2.1. Flat background: the Banks--Peskin differential forms, the
Feigin--Frenkel--Garland--Zukerman fromalism, the Siegel string fields and the
Kato--Ogawa BRST--operator
\endsubhead

Let us introduce the basic notions [1,58,59]: a closed string is an arbitrary
contour $C$ in D--dimensional space $\Bbb R^D$ (it is usually supposed that
$\Bbb R^D$ possesses a Minkowsky metric and the transition to the Euclidean
one is done later). a collection of such contours will be denoted by $Q_0$.
One might define a classical string field as a magnitude on $Q_0$. But this is
not convenient. the fact is that $Q_0$ is not a smooth manifold. It has
singularities. To define a string field one should specify its behaviour near
singularity. It is opportune to make occasion of the possibility of a
representation of $Q_0$ as an orbifold to achieve our object. Indeed, let us
parametrise a contour $C$ by a function $x^\mu(s):[-\pi,\pi]\mapsto\Bbb R^D$.
We denote the set of such functions by $Q$, which admits an action of the
group $\Diff_+(\circle)$. The initial space $Q_0$ is a quotient
$Q/\Diff_+(\circle)$. So a classical scalar string field maybe defined as a
magnitude on $Q$. In a parametrisation $x^\mu(s)$ a string field has the form
$\Phi(x^\mu)$. The independence of $\Phi$ on a choice of parametrisation is
called reparametrisation invariance. Let us find a law of string--field
transformation under reparametrisations, supposing that it is determined by
the law of transformations of the first quantised string coordinates
$x^\mu_n$: $x^\mu(s)=\real\sum x^\mu_nz^{-n}$ ($s=\exp(is)$). This law should
be derived from the first quantised closed string action. However, the action
can not be received in a unique way, because it should be obtained only from
some physical assumptions on the behaviour of magnitudes on $Q_0$ instead of
$Q$. Hence, the action will contain a gauge parameter. The choice of the first
quantised closed string action is one of A.M.Polyakov [60]:
$$ S(x^\mu)=\int g(z,\bar z)\partial_zx^\mu(z)\partial_{\bar z}\bar x^\mu(z)\,
dzd\bar z. \tag 2A$$
Here we assume the holomorphy of the coordinates $x^\mu(z)$ on a world surface
of a string in view of the Hamiltonian character of its evolution. Let us
interpret the metric $g(z,\bar z)$ as a gauge parameter. In the gauge
$g(z,\bar z)=1$ one gets the next expression for the action
$$S(x^\mu)=\int_{D_+}\partial_zx^\mu(z)\partial_{\bar z}\bar x^\mu(z)\,
dzd\bar z=\sum_{n>0}nx^\mu_n\bar x^\mu_n,\tag 2B$$
where $D_+=\{z:|z|\le 1\}$. The first quantisation of a string with such
action is described in [1,58,59]. In the space of states of a first quantised
string (functions of the infinite number of variables $x^\mu_n$ obeying the
equation $(\pd{x^\mu_0}-e_\mu)\Phi=0$) the Virasoro algebra generators
$L_{-p}$ with $p\ge 0$ naturally act:
$$L_{-p}=e_\mu px^\mu_p+\frac12\sum_{k=1}^{p-1}k(p-k)x^\mu_kx^\mu_{p-k}+
\sum_{k\ge 1}(k+p)x^\mu_{k+p}\pd{x^\mu_k}.$$
After a transition to the second quantisation the Fock space $F(Q^*)$ dual to
the space of the first quantised string states is interpreted as a
configuration space for the second quantised string.

However, the action (2B) is not gauge invariant. The mechanism accounting for
the noninvariance of the gauge was proposed in the papers [61,62,19,63].
Accordingly to [19] expression (2A) depends on a choice of a complex structure
on $Q^*$. The space of such structures is isomorphic to
$M(\Vir)=\Diff_+(\circle)/\circle$. An element of the space $M(\Vir)$, an
univalent function $f$, generates global transformations of the fields
$\Phi(x^\mu)$ preserving the family of Polyakov actions. In such case one
should postulate a locality of transformations. the relation between the gauge
parameters $g(z,\bar z)$ and $f(z)$ has the form $g(z,\bar z)dx^\mu d\bar
x^\mu=dx^\mu(f)d\bar x^\mu(f)$. Let us now define the configuration space of
the closed string field as a space of magnitudes $\Phi(x^\mu,f)$. The
decomposition of a field by coefficients $c_k$ independent of $x^\mu_n$
determines a multicomponent string field $\Phi(x^\mu,f)=\sum_\xi
c_1^{\xi_1}\ldots c_n^{\xi_n}\Phi_\xi(x^\mu)$. The new action has the form
[63]
$$ S(x^\mu,f)=\int dx^\mu(f)d\bar x^\mu(f)+U_{h,c},\tag 2C$$
where $U_{h,c}(f)$ are the K\"hler potentials on $M(\Vir)$ and $\int
dx^\mu d\bar x^\mu$ is the K\"ahler potential on $Q^*$. the space dual to the
space of string fields consists of all holomorphic sections of the Hermitean
bundle $E_{h,c}$, the action of the Virasoro algebra in $\Cal O(E_{h,c})$ has
the form
$$\allowdisplaybreaks\align
L_p=&\pd{c_p}+\sum_{k\ge 1}(k+1)c_k\pd{c_{k+p}}\quad (p>0),\\
L_0=&\sum_{k\ge 1}kc_k\pd{c_k}+\sum_{k\ge 1}kx^\mu_k\pd{x^\mu_k}+h,\\
L_{-1}=&\sum_{k\ge 1}((k+2)c_{k+1}-2c_1c_k)\pd{c_k}+2c_1\sum_{k\ge
1}kx_k^\mu\pd{x^\mu_k}+\\&\sum_{k\ge 1}(k+1)x^\mu_{k+1}\pd{x^\mu_k}+
2hc_1+e_\mu x^\mu_1,\\
L_{-2}=&\sum_{k\ge 1}((k+3)c_{k+2}-(4c_2-c_1^2)c_k-b_k(c_1,\ldots
c_{k+2}))\pd{c_k}+(4c_2-c^2_1)\sum_{k\ge 1}kx_k^\mu\pd{x^\mu_k}+\\
&3c_1\sum_{k\ge 1} x^\mu_{k+1}\pd{x^\mu_k}+\sum_{k\ge
1}(k+2)x^\mu_{k+2}\pd{x^\mu_k}+\\
&+\frac{x^2_1}2+h(4c_2-c_1^2)+\frac{c}2(c_2-c_1^2)+3e_\mu c_1x^\mu_1+
2e_\mu x^\mu_2,\\
L_{-n}=&\frac1{(n-2)!}\ad^{n-2}L_{-1}\cdot L_{-2}\quad (n>2).
\endalign
$$
The dual space to $\Cal O(E^*_{h,c})$ (the Fock space which is a semidirect
product of the Fock space over the flag manifold for the Virasoro--bott group
considered in the first chapter and the standard (flat) Fock space) is the
configuration space for the closed string--field theory (without ghosts). The
action of the Virasoro algebra in it has the form
$$\allowdisplaybreaks\align
L_{-p}=&\,c_p+\sum_{k\ge 1}(k+1)c_{k+p}\pd{c_k}\quad (p>0),\\
L_0=&\sum_{k\ge 1}kc_k\pd{c_k}+\sum_{k\ge 1}kx^\mu_k\pd{x^\mu_k}+h,\\
L_1=&\sum_{k\ge 1}c_k((k+2)\pd{c_{k+1}}-2\pd{c_1}\pd{c_k})+2\sum_{k\ge
1}kx^\mu_k\pd{c_1}\pd{x^\mu_k}+\\
&\sum_{k\ge 1}(k+1)x^\mu_k\pd{x^\mu_{k+1}}+2h\pd{c_1}+
e^\mu\pd{x^\mu_1},\\
L_2=&\sum_{k\ge
1}c_k((k+3)\pd{c_{k+2}}-(4\pd{c_2}-(\pd{c_1})^2)\pd{c_k}-
b_k(\pd{c_1},\ldots\pd{c_{k+2}}))+\tag 3A\\&\sum_{k\ge
1}kx^\mu_k\pd{x^\mu_k}(\pd{c_2}-(\pd{c_1})^2)+3\sum_{k\ge
1}(k+1)x^\mu_k\pd{c_1}\pd{x^\mu_{k+1}}+\\
&\sum_{k\ge 1}(k+2)x^\mu_k\pd{x^\mu_{k+2}}+\frac12\frac{\partial^2}{\partial
x^2_1}+h(4\pd{c_2}-(\pd{c_1})^2)+\frac{c}2(\pd{c_2}-
(\pd{c_1})^2)+\\
&3e^\mu\pd{x^\mu_1}\pd{c_1}+2e^\mu\pd{x^\mu_2},\\
L_n=&\frac{(-1)^n}{(n-2)!}\ad^{n-2}L_1\cdot L_2\quad (n>2).
\endalign
$$
A metric on the configuration space has the form
$$(\Phi,\bar\Psi)=(\Phi|\exp(S)|\Psi)=:
\int\exp(-S(x^\mu,f))\Phi(x^\mu,f)\Psi(x^\mu,f)\,DXD\bar XDfD\bar f.$$
The area of integration in the last integral is determined by the univalency
conditions for the function $f$. It should be mentioned that
$\exp(S)=\exp(\int dx^\mu(f)d\bar x^\mu(f))\cdot K_{h,c}(f)$. The second term
in the formula is the Bergman kernfunction on the space of the univalent
functions (after the identification of this space with the universal
Teichm\"uller space the Bergman kernfunction will coincide with the Polyakov
measure [64,65] as it was shown by A.Morozov and A.Rosly in the paper [66],
namely,
$K_{h,c}(f)=\exp(U_{h,c}(f))=(\operatorname{cap}(f(D_+)))^h\cdot
\det^{-c}(1-Z_f\bar Z_f)$, where $\operatorname{cap}(f(D_+))$ is the conformal
capacity of the domain $f(D_+)$.

\define\BP{\operatorname{BP}}
\define\SI{\operatorname{SI}}
\define\ghost{\operatorname{ghost}}
\define\vac{\operatorname{vac}}
To quantise a field $\Phi(x^\mu,f)$ it is necessary to supplement the support
manifold by the anticommuting Faddeev--Popov ghosts. They are transformed
accordingly to the adjoint representation of the constraint algebra
$\CVect(\circle)$ (usually parallel with ghosts one consider also elements of
dual space, which transform accordingly to the coadjoint representation of the
constraint algebra). It is convenient to think ghosts as vector fields on
$M(\Vir)\ltimes Q^*$ tangent to the constraint foliation. The dual family of
differential froms (antighosts) $\xi_p$ with the law of transformation
$L_p\xi_q=-(q+2p)\xi_{p+q}$ generates the space of the Banks--Peskin string
differential forms [61]. So the string differential forms depend on the string
coordinates $x^\mu_n$ in the external space, the internal degrees of freedom
of the string $c_k$ and Faddeev--Popov antighosts $\xi_p$. The subspace of
such forms in $\Omega(M(\Vir)\ltimes Q^*)$ will be denoted as $\Omega_{\BP}$.
Let us introduce according to Feigin, Frenkel, Garland and Zukerman [67,68]
the vacuum vector corresponding to the filled ghost Dirac sea relative to the
subalgebra $<L_p,p>0>$ as $\vac=\xi_1\wedge\xi_2\wedge\xi_3\wedge\ldots$.
Consider now the space $\Omega^{\SI}_{\BP}$ of semi--infinite string
differential forms (the strict definition of the semi--infinite form maybe
found in [67,68]), and also $\Omega^{\SI}_{\BP}(E_{h,c})=\Omega^{\SI}_{\BP}
\otimes_{\Cal O(M(\Vir)\ltimes Q^*)}\Cal O(E_{h,c})$. The Virasoro algebra
acts in $\Omega^{|SI}_{\BP}$ with the central charge $c=-26$ and the value of
$L_0$ on the vacuum $\vac$ being equal to $h-1$. The formulas for the action
have the form
$$L'_{-p}=L_{-p}+L^{\ghost}_{-p},\quad
L^{\ghost}_{-p}=\sum_q(p-2q)\xi_{q-p}\pd{\xi_q}.\tag 3B$$

Let us introduce the Kato--Ogawa BRST--operator [69] as a partial differential
along the constraints $Q=\sum_pL'_{-p}\xi_p$. A request for the absence of the
conformal anomaly (the nilpotency of the BRST--operator $Q$: $Q^2=0$) picks
out the value of the parameter $c$: $c=26$.

\definition{\rm DEFINITION [62,63]} The Siegel string field is an element of
the space $\Omega^{\SI}_{\BP}(E^*_{h,c})^*$ dual to the space of the
Banks--Peskin differential forms.
\enddefinition

Therefore, the space of Siegel string fields is a product of the Fock space of
the bundle $E_{h,c}$ over $M(\Vir)\ltimes Q^*$ and the space of the
semi--infinite ghost forms. the operator $Q^*$ conjugate to $Q$ (which is also
called by the Kato--Ogawa BRST--operator) has the form
$Q^*=\sum_{p,q}(L_{-p}\xi^*_p-\frac12(p-q)\xi^*_p\xi^*_q)\pd{\xi^*_{-(p+q)}}$,
where $(\xi^*_p,\xi_q)=-\delta(p+q)$. The scalar product in
$\Omega^{\SI}_{\BP}(E^*_{h,c})^*$ maybe defined as
$$(\Phi,\bar\Psi)=\int\exp(-S)\Phi^{(\alpha)}\bar\Psi^{(\beta)}\,DXD\bar
XDfD\bar f (\xi^*_{(\alpha)},\xi^*_{(\beta)}),$$
where $\Phi=\Phi^{(\alpha)}\xi^*_{(\alpha)}$,
$\Psi=\Psi^{(\beta)}\xi^*_{(\beta)}$ and $(\xi^*_{(\alpha)},\xi^*_{(\beta)})$
is the scalar product in the space of the semi--infinite ghost forms [67,68].

At the end of this article we shall give a cohomological interpretation of the
Nambu--Goto action [1,58,59], accordingly to [70]. For such purpose let us
consider a representation of the Virasoro algebra in the space $V_D$ of
functions of the variables $c_1,\ldots c_n,\ldots x^\mu_1,\ldots
x^\mu_k,\ldots$ by the differential operators
$$\allowdisplaybreaks\align
L_p=&\pd{c_p}+\sum_{k\ge 1}(k+1)c_k\pd{c_{k+p}}\quad (p>0),\\
L_0=&\sum_{k\ge 1}kc_k\pd{c_k}+\sum_{k\ge 1}kx^\mu_k\pd{x^\mu_k},\\
L_{-1}=&\sum_{k\ge 1}((k+2)c_{k+1}-2c_1c_k)\pd{c_k}+2c_1\sum_{k\ge
1}kx_k^\mu\pd{x^\mu_k}+\\&\sum_{k\ge 1}(k+1)x^\mu_{k+1}\pd{x^\mu_k},\tag 4\\
L_{-2}=&\sum_{k\ge 1}((k+3)c_{k+2}-(4c_2-c_1^2)c_k-b_k(c_1,\ldots
c_{k+2}))\pd{c_k}+(4c_2-c^2_1)\sum_{k\ge 1}kx_k^\mu\pd{x^\mu_k}+\\
&3c_1\sum_{k\ge 1} x^\mu_{k+1}\pd{x^\mu_k}+\sum_{k\ge
1}(k+2)x^\mu_{k+2}\pd{x^\mu_k},\\
L_{-n}=&\frac1{(n-2)!}\ad^{n-2}L_{-1}\cdot L_{-2}\quad (n>2).
\endalign
$$

\define\NG{\operatorname{NG}}
\define\vect{\operatorname{vect}}
\proclaim{\rm THEOREM 6 [70]} $H^1(\Cvir,V_D)=\operatorname{Sym}(\Bbb C,D)+
\Bbb C^D+\Bbb C_c$. The matrix cohomology class is determined by the
Nambu--Goto action $S_{\NG}(G_{ab})=\sum_n nG_{ab}\tilde x^a_n\overline{\tilde
x}^a_n$:
$$\align & L'_{-p}=L_{-p}+F^{\NG}_{-p},\\
&F^{\NG}_{-p}=\frac12\sum_{m+n=p}nmG_{ab}\tilde x^a_m\tilde x^b_n,
\text{ \sl if } p>0\text{ \sl and } 0 \text{ \sl otherwise.}
\endalign
$$
The vector class has the form
$$\align & L'_{-p}=L_{-p}+F^{\vect}_{-p},\\
&F^{\vect}_{-p}=b_ap(p-1)\tilde x^a_p,\text{ \sl if }
p>0\text{ \sl and } 0 \text{ \sl otherwise.}
\endalign
$$
The class corresponding to parameter $c$ is defined by the formulas written in
the first chapter. Here $\tilde x^\mu_n=(x^\mu(f))_n$.
\endproclaim

The statement of the theorem maybe obtained by the straightforward
calculations.

\subhead 2.2. Curved background [71] \endsubhead

It should be mentioned that the Banks--Peskin differential forms have no
a geometric meaning; on the contrary, the string fields are the correctly
defined objects on the background. Let $v^j_{k_1\ldots k_D}=(x^1)^{k_1}\ldots
(x^D)^{k_D}$ be a vector field on the background then the natural action of
the field $v^j_{k_1\ldots k_D}$ in the space of string fields has the form
[71]
$$\align T(v^j_{k_1\ldots k_D}=&\sum_{\sum m_{ij}=M}\tilde x^1_{m_{11}}\ldots
\tilde x^1_{m_{1k_1}}\ldots\tilde x^D_{m_{D1}}\ldots\tilde x^D_{m{Dk_D}}
\pd{\tilde x^j_M},\\
&\tilde x^i_n=(x^i(f))_n.
\endalign
$$
If $G_{ab}=0$ then $[T(v^j_{k_1\ldots k_D}),L_p]=0$, where $L_p$ is defined by
formulas dual to (4).

Let us fix a non--constant metric $G_{\alpha\beta}$ on the background. The
Laplace operator for $G_{\alpha\beta}$ will have the form
$$\Delta(G_{\alpha\beta})=G^{ij}(x^1,\ldots x^D)\frac{\partial^2}{\partial
x^i\partial x^j}+H^i(x^1,\ldots x^D)\pd{x^i},$$
where
$$H^i=G^{-\frac12}\pd{x^j}(G^{\frac12}\cdot G^{ij}).$$

Let us now consider the operators $[\Delta(G_{\alpha\beta})]_p$ defined as

$$\align
[\Delta(G_{\alpha\beta})]_p=&\sum_{b+c-|a|=p} f(a,b,c)\tilde
G^{ij}(x^1_{a_{11}}\ldots x^1_{a_{1m_1}}\ldots x^D_{a_{D1}}\ldots
x^D_{a_{Dm_D}})\frac{\partial^2}{\partial x^i_b\partial^j_c}+\\
&\frac16\sum_{b-|a|=p}(b^3-b)\tilde H^i(x^1_{a_{11}}\ldots x^1_{a_{1m_1}}\ldots
x^D_{a_{D1}}\ldots
x^D_{a_{Dm_D}})\pd{x^i_b},
\endalign
$$
where $\tilde G^{ij}$ and $\tilde H^i$ are the polylinearisations of $G^{ij}$
and $H^i$, the function $f(a,b,c)$ is defined as
$$f(a,b,c)=\sum_{A'\sqcup A''}(b-\sum_{i\in A'}a_i)(c-\sum_{i\in A''}a_i).$$

\proclaim{\rm THEOREM 7 [71]}
The operators
$$ G_{\alpha\beta}(L_p)=L_p\text{ \sl of (4) if } p>0\text{ \sl and } L_p
\text{ \sl of (4) }+\frac12[\Delta(G_{\alpha\beta)})]_{-p}(\tilde x^i_n)\text{
\sl if } p\le 0,$$
where $\tilde x^i_n=(x^i(f))_n$, define an action of the Virasoro algebra in
the space of string fields. The correspondence $G_{\alpha\beta}\longrightarrow
G_{\alpha\beta}(\cdot)$ is generally covariant.
\endproclaim

The corresponding BRST--operator $G_{\alpha\beta}(Q)$ has the form
$$G_{\alpha\beta}(Q)=\sum(L_{-p}\xi^*_p-\frac12(p-q)\xi^*_p\xi^*_q
\pd{\xi^*_{-p-q}}).$$ The BRST--operator is nilpotent if $c=26$.

\subhead 2.3. The Bowick--Rajeev formalism \endsubhead

The Bowick--Rajeev formalism [19] describes a separation of variables,
characterising external and internal degrees of freedom of a string in quantum
field theory. Let us present the Bowick--Rajeev formalism for the flat
background following to [72]. Let us consider the space
$\Omega^{\SI}_{\BP}(E_{h,c})$ of the semi--infinite Banks--Peskin differential
forms. The action of the Virasoro algebra has the form
$$\allowdisplaybreaks\align
L_p=&\pd{c_p}+\sum_{k\ge
1}(k+1)c_k\pd{c_{k+p}}-\sum_k(p+2k)\xi_{k+p}\pd{\xi_k}\quad (p>0),\\
L_0=&\sum_{k\ge 1}kc_k\pd{c_k}+\sum_{k\ge
1}kx^\mu_k\pd{x^\mu_k}-2\sum_kk\xi_k\pd{\xi_k}+h,\\
L_{-1}=&\sum_{k\ge 1}((k+2)c_{k+1}-2c_1c_k)\pd{c_k}+2c_1\sum_{k\ge
1}kx_k^\mu\pd{x^\mu_k}+\\&\sum_{k\ge
1}(k+1)x^\mu_{k+1}\pd{x^\mu_k}+\sum_k(1-2k)\xi_{k-1}\pd{\xi_k}+
2hc_1+e_\mu x^\mu_1,\\
L_{-2}=&\sum_{k\ge 1}((k+3)c_{k+2}-(4c_2-c_1^2)c_k-b_k(c_1,\ldots
c_{k+2}))\pd{c_k}+(4c_2-c^2_1)\sum_{k\ge 1}kx_k^\mu\pd{x^\mu_k}+\\
&3c_1\sum_{k\ge 1} x^\mu_{k+1}\pd{x^\mu_k}+\sum_{k\ge
1}(k+2)x^\mu_{k+2}\pd{x^\mu_k}+\\
&+\sum_k2(1-k)\xi_{k-2}\pd{\xi_k}+\frac{x^2_1}2+h(4c_2-c_1^2)+
\frac{c}2(c_2-c_1^2)+3e_\mu c_1x^\mu_1+2e_\mu x^\mu_2,\\
L_{-n}=&\frac1{(n-2)!}\ad^{n-2}L_{-1}\cdot L_{-2}\quad (n>2).
\endalign
$$

\define\FG{\operatorname{FG}}
The space $\Omega^{\SI}_{\BP}(E_{h,c})$ maybe considered as a space of
sections of some vector bundle over $M(\Vir)$, a fiber $H_f(h,c)$ of which is
isomorphic to $H(h,c)=F(Q)\otimes\bigwedge^{\SI}(\operatorname{\Bbb
CVect}^*(\circle))$. We shall denote this bundle by $\FG_{h,c}(M(\Vir))$, the
variables $x^\mu_n$, $c_k$ define its trivialisation.

A gauge of a Banks--Peskin string differential form is a relation
$c_k=c_k(x^\mu)$, and a gauge--fixing projector $P$ is the operator
$P:\Omega^{\SI}_{\BP}(E_{h,c})\mapsto H(h,c)$ determined by the formula
$$P\Phi=\left.\Phi\right|_{c_k=c_k(x^\mu)}.$$ Let $f(z)=z+c_1^\circ
z^2+c_2^\circ z^3+c_3^\circ z^4+\ldots$ be an arbitrary univalent function,
the $f$--gauge is the gauge defined by the relation $c_k=c^\circ_k$.
The gauge--fixing projector has the form
$$P_f\Phi=\left.\Phi\right|_{c_k=c_k^\circ}.$$

As it was shown in [72] there exists an imbedding
$I_f:H(h,c)\mapsto\Omega^{\SI}_{\BP}(E_{h,c})$ such that (1) $\Cal
L_+(f)I_f=0$, where $\Cal L_+(f)$ is a set of positive--frequency generators
of the Virasoro algebra determined by a choice of a gauge parameter value
(i.e. if $f=g\cdot f_0$, $f_0(z)=z$, $g\!\in\!\Diff_+(\circle)$, then $\Cal
L_+(f)=g\cdot\span(L_n,n>0)$); (2) $I_f\Phi$ in gauge $f$ equals to $\Phi$,
i.e. $P_fI_f=\id$.

The family $I=\{I_f\}$ defines an imbedding
$$I:\FG_{h,c}(M(\Vir))\mapsto M(\Vir)\times\Omega^{\SI}_{\BP}(E_{h,c}).$$
Define a connection $\nabla$ in $\FG_{h,c}(M(\Vir))$ such as
$$\nabla_X\varphi_f=P_f(X(I_f\varphi_f))$$
or
$$\nabla_X\varphi_f=\lim_{t\to
0}t^{-1}(P_fI_{f+tXf}\varphi_{f+tXf}-\varphi_f).$$

The connection  $\nabla$ maybe considered as an infinite dimensional analogue
of the Gauss--Manin connection. the connection is not always flat. the
condition of an absence of a curvature put a restrictions on values of
parameters $d$, $c$, $h$, $e_\mu$.

\proclaim{\rm THEOREM 8 [72]} The curvature tensor of the connection $\nabla$
in the bundle $\FG_{h,c}(M(\Vir))$ is equal to

$$R_{n,m}=((c-D)\frac{n^3-n}{12}+(2h-e_\mu e^\mu))\delta(n+m).$$
\endproclaim

\definition{\rm DEFINITION} The covariantly constant section of the bundle
$\FG_{h,c}(M(\Vir))$ is called the Bowick--Rajeev vacuum.
\enddefinition

The Bowick--Rajeev vacuum exists if and only if $D=c$, $h=e^2/2$.

Define following [72] the family of parings
$$B_{f_1,f_2}(\cdot,\cdot):((\Omega^{\SI}_{\BP}(E^*_{h,c}))^*)^{\otimes
2}\mapsto\Bbb C$$
as
$$ B_{f_1,f_2}(\Phi,\Psi)=(P^*_{f_1}I^*_{f_1}\Phi,P^*_{f_2}I^*_{f_2}\Psi).$$
The pairings $B_{f_1,f_2}(\cdot,\cdot)$ play a role of the
Kostant--Blattner--Sternberg pairings [72]. Indeed, as it was shown in [72],
if $c>1$, $h>0$ the connection $\nabla$ maybe defined in the following way
$$(\nabla^*_X\Phi=0)\Leftrightarrow(B_{f+tXf,f}(\Phi,\Psi)-B_{f,f}(\Phi,\Psi)=
o(t),\forall\Psi).$$

\define\GI{\operatorname{GI}}
\definition{\rm DEFINITION} A covariantly constant with respect to $\nabla^*$
element of the space $\Omega^{\SI}_{\BP}(E^*_{h,c})^*$ of the string fields is
called a gauge--invariant string field; the space of such fields will be
denoted by $\Omega^{\SI}_{\BP}(E^*_{h,c})^*_{\GI}$.
\enddefinition

the space of the gauge--invariant string fields is dual to the space of the
Bowick--Rajeev vacuums.

\proclaim{\rm THEOREM 9A [72]}
The space of the gauge--invariant string fields
$\Omega^{\SI}_{\BP}(E^*_{h,c})^*_{\GI}$ is invariant under the Virasoro
algebra action. Let us identify the space
$\Omega^{\SI}_{\BP}(E^*_{h,c})^*_{\GI}$ with $H_f(h,c)$ by the operator $I_f$.
In $H_f(h,c)$ the Virasoro algebra generators act by operators $T_f(L_k)$. If
$f=f_0$ then
$$T_{f_0}(L_k)=L^V_k+L_k^{\ghost},$$
where $L^V_k$ are the generators in the Virasoro representation
$$\allowdisplaybreaks\align
L^V_p=&\sum_{k\ge 1}(k+p)x^\mu_k\pd{x^\mu_{k+p}}+\frac12\sum_{k=1}^{p-1}k(p-k)
\frac{\partial^2}{\partial x^\mu_k\partial x^\mu_{p-k}}+pe^\mu\pd{x^\mu_p},\\
L^V_0=&\sum_{k\ge 1} kx^\mu_k\pd{x^\mu_k}+\frac{e^2}2,\\
L^V_{-p}&=\sum_{k\ge 1}kx^\mu_{k+p}\pd{x^\mu_k}+\frac12\sum_{k=1}^{p-1}x^\mu_k
x^\mu_{p-k}+e_\mu x^\mu_p,
\endalign
$$
and $L_k^{\ghost}$ are the generators of the Virasoro algebra in the ghost
space
$$L_p^{\ghost}=\sum_q(q-p)\xi^*_q\pd{\xi^*_{p+q}}.$$
\endproclaim

\proclaim{\rm THEOREM 9B [72]} The action of the Virasoro algebra in $(\Cal
O(\FG^*_{h,c}(M(\Vir))))^*$ has the form
$$\left. L_K\right|_f={\left.\nabla(L_k)\right|_f}^*+T_f(L_k),$$
$D=c$, $e^2=2h$.
\endproclaim

It should be mentioned that the space $\Omega^{\SI}_{\BP}(E^*_{h,c})^*$ of the
gause--invariant string fields is $Q^*$--invariant, where $Q^*$ is the
conjugate Kato--Ogawa BRST--operator [69]. Let us define a fiber
BRST--operator $Q^*_0$ as $\left.
Q^*\right|_{\Omega^{\SI}_{\BP}(E_{h,c}^*)^*_{\GI}}$, where
$\Omega^{\SI}_{\BP}(E^*_{h,c})^*_{\GI}$ is identified with $H(h,c)$ by
$I^*_f$.

\proclaim{\rm THEOREM 9C [72]} $Q^*=D_\nabla+Q^*_0$. The fiber BRST--operator
$Q^*_0$ under the initial fiber has the form $Q^*_0=\sum(L^V_{-p}+\frac12
L^{\ghost}_{-p})\xi^*_p$.
\endproclaim

The Bowick--Rajeev formalism maybe expanded on a curved background: the gauges
of the
Banks--Peskin string differential forms, the projectors, which fix these
gauges, the Connection $\nabla$, the Bowick--Rajeev vacua, the
gauge--invariant string fields, the fiber BRST--operators can be introduced
analogously to the flat case. the main problem in the case problem, which is
not solved completely yet, is to characterise the values of parameters of the
curved background for which the Bowick--Rajeev vacua and the gauge--invariant
string fields exist. the differential equations on such parameters providing
the presence of the gauge--invariant string fields are called the string
Einshtein equations.

\Refs
\roster
\item"[1]" Morozov A.Yu., What is the string theory? Uspekhi Fiz. Nauk 168:2
(1992), 83-176 [in Russian].
\item"[2]" Green M.B., Schwarz J.H., Witten E., {\it Superstring theory}.
Cambridge Univ. Press, Cambridge, 1988.
\item"[3]" Juriev D.V., Quantum conformal field theory as infinite dimensional
noncommutative geometry. Uspekhi Matem. nauk 46:4 (1991), 115-138 [in
Russian].
\item"[4]" Neretin Yu.A., Holomorphic extensions of the representations of the
group of diffeomorphisms of a circle. Matem. Sb. 180:5 (1989), 635-657 [in
Russian].
\item"[5]" Neretin Yu.A., Infinite dimensional groups, their mantles, trains
and representations. Adv. Soviet Math. 2 (1991), 103-172.
\item"[6]" Kirillov A.A., A K\"ahler structure on K--orbits of the group of
diffeomorphisms of a circle. Funkt. anl. i ego prilozh. 21:1 91987), 42-45 [in
Russian].
\item"[7]" Kirillov A.A., The flag manifold for the Virasoro--Bott group. In
{\it Infinite dimensional Lie lagberas and quantum field theory (Varna,
1987)}, World Scientific, Teaneck, NJ, 1988, 73-77.
\item"[8]" Kirillov A.A., Juriev D.V., The K\"ahler geometry of the infinite
dimensional homogeneous space $M=\Diff_+(\circle)/\Rot(\circle)$. Funkt. anal.
i ego prilozh. 21:4 (1987), 35-46 [in Russian].
\item"[9]" Segal G., Unitary representations of some infinite dimensional
groups. Commun. Math. Phys. 80 (1981), 301-342.
\item"[10]" Kirillov A.A., Infinite dimensional Lie groups, their orbits,
invariants and representations. Lect. Notes Math. 970 (1982), 101-123.
\item"[11]" Witten E., Coadjoint orbits of the Virasoro group. Commun. Math.
Phys. 114 (1988), 1-53.
\item"[12]" Goluzin G.M., {\it Geometric theory of functions of a complex
variable}. Moscow, "Nauka", 1968 [in Russian].
\item"[13]" Jenkins J., {\it Univalent functions and conformal mapping}.
Springer, 1958.
\item"[14]" Duren P.L., {\it Univalent functions}. Springer, 1983.
\item"[15]" Juriev D.V., On the univalency radius of a regular function.
Funkt. anal. i ego prilozh. 24:1 (1990), 90-91 [in Russian].
\item"[16]" Juriev D.V., On the determining of the univalency radius of a
regular function by its Taylor coefficients. Matem. Sb. 183:1 (1992), 45-54
[in Russian].
\item"[17]" Kirillov A.A., Juriev D.V., Representations of Virasoro algebra by
orbit method. J.Geom. Phys. 5 (1988), 351-363 [reprinted in {\it Geometry and
physics. Essays in honour of I.M.Gel\-fand}. Eds. S.G.Gindikin and I.M.Singer.
Elsevier Sci. Publ. and Pitagora Editrice, 1991].
\item"[18]" Juriev D., The vocabulary of geometry and harmonic analysis on the
infinite dimensional manifold $\Diff_+(\circle)/\circle$. Adv. Soviet Math. 2
(1991), 233-247.
\item"[19]" Bowick M.J., Rajeev S.G., The holomorphic geometry of closed
bosonic string theory and $\Diff(\circle)/\circle$. Nucl. Phys. B293 (1987),
348-384.
\item"[20]" Fuchs D.B., {\it Cohomology of infinite dimensional Lie algebras}.
Moscow, "Nauka", 1984 [in Russian].
\item"[21]" Manin Yu.I., {\it Cubic forms: arithmetics, algebra, geometry}.
Moscow, "Nauka", 1972 [in Russian].
\item"[22]" Belousov V.D., {\it Foundations of the theory of quasigroups and
loops}. Moscow, "Nauka", 1967 [in Russian].
\item"[23]" Sabinin L.V., Methods of nonassociative algebra in differential
geometry. Suppl. to the Russian transl. of S.Kobayashi and K.Nomizu, {\it
Foundations of differential geometry}. V.1, Moscow, "Nauka", 1981 [in
Russian].
\item"[24]" Sabinin L.V., Mikheev P.O., Quasigroups and differential geometry.
Problems of geometry 20, Moscow, VINITI, 1988 [in Russian].
\item"[25]" Karasev M.V., Maslov V.P., {\it Nonlinear Poisson brackets:
geometry and quantization}. Moscow, "Nauka", 1991 [in Russian].
\item"[26]" Rund H., {\it The differential geometry of Finsler spaces}.
Springer, 1959.
\item"[27]" Asanov G.S., {\it Finslerian extensions of General Relativity}.
Dordrecht, 1984.
\item"[28]" Juriev D.V., Non--Euclidean geometry of mirrors and
prequantization on the homogeneous K\"ahler manifold
$M=\Diff_+(\circle)/\circle$. Uspekhi Matem. Nauk 43:2 (1988), 159-160 [in
Russian].
\item"[29]" Juriev D., An infinite dimensional geometry of the universal
deformation of a complex disk. Russian J. Math. Phys. 2 (1994);
funct-an/9401003.
\item"[30]" Hua Loo--Keng, Harmonic analysis in classical domains. Moscow,
1959 [in Russian].
\item"[31]" Sato M., Sato Y., Soliton equations as dynamical system on
infinite dimensional Grassmann manifold. In {\it Nonlinear partial
differential equations in applied science}, Amsterdam, 1983, 259-271.
\item"[32]" Segal G., Wilson G., Loop groups and equations of KdV type. Publ.
Math. IHES 61, 5-65.
\item"[33]" Pressley A., Segal G., {\it Loop groups}, Oxford, 1988.
\item"[34]" Grunsky H., Koeffizientenbedingungen f\"ur schlicht abbildende
meromorphe Funktionen. math. Z. 45 (1939), 29-61.
\item"[35]" Krichever I.M., Methods of algebraic geometry in the theory of
nonlinear equations. Uspekhi Matem. Nauk 32:6 (1977), 183-208 [in Russian].
\item"[36]" Milin I.M., {\it Univalent functions and orthonormal systems}.
Moscow, "Nauka", 1971 [in Russian].
\item"[37]" Gershkovich V., Vershik A., Nonholonomic manifolds and nilpotent
analysis. J. Geom. Phys. 5 (1988), 407-452 [reprinted in {\it Geometry and
physics. essays in honour of I.M.Gel\-fand}. Eds. S.G.Gindikin and I.M.Singer.
Elsevier Sci. Publ. and Pitagora Editrice, 1991].
\item"[38]" Kac V.G., {\it Infinite dimensional Lie algebras}. Cambridge,
Cambridge Univ. Press, 1990.
\item"[39]" Feigin B.L., Fuchs D.B., Representations of the Virasoro algebra.
In {\it Representations of infinite dimensional Lie algebras}. Gordon and
Beach, 1991.
\item"[40]" Tsuchiya A., Kanie Y., Fock space representations of the Virasoro
algebra. Intertwinning operators. Publ. RIMS 22 (1986), 259-327.
\item"[41]" Juriev D.V., A model of the Verma modules over the Virasoro
algebra. Algebra i analiz 2:2 (1990), 209-226 [in Russian].
\item"[42]" Gelfand I.M., Gindikin S.G., Complex manifolds whose skeletons are
the semisimple real Lie groups and analytic discrete series of
representations. Funkt. anal. i ego prilozh. 11:4 (1977), 19-27 [in Russian].
\item"[43]" Olshanskii G.I., Invariant cones in Lie algberas, Lie semigroups
and holomorphic discrete series. Funkt. anal. i ego prilozh. 15:4 91981),
53-66 [in Russian].
\item"[44]" Bychkov S.A., Plotnikov S.V., Juriev D.V., Skladen' of the Verma
modules over the Lie algebra $\sLtwo$ and hidden $\sLthree$--symmetries in the
quantum projective field theory. Uspekhi Matem. Nauk 47:3 (1992), 153 [in
Russian].
\item"[45]" Tumanov A.E., Geometry of CR--manifolds. Current Probl. Math.,
Fundam. Directions 9 (1986), 225-256 [in Russian].
\item"[46]" Alekseev A.Yu., Shatashvili S.L., Path quantization of the
coadjoint orbits of the Virasoro group and 2d gravity. Nucl. Phys. B323
(1989), 719-733.
\item"[47]" Alekseev A.Yu., Shatashvili S.L., From geometric quantization to
conformal field theory. Commun. Math. Phys. 128 (1990), 197-212.
\item"[48]" Ahlfors L., Bers L., {\it Spaces of Riemann surfaces and
quasiconformal mappings}. Moscow, 1961 [in Russian].
\item"[49]" Lehto O., {\it Univalent functions and Teichm\"uller spaces}.
Springer, 1986.
\item"[50]" Alvarez--Gaume L., Gomez C., Reina C., Loop groups, grassmannians
and string theory. Phys. Lett. B190 (1987), 55-62.
\item"[51]" Morozov A.Yu., Perturbative methods in string theory. Elem. Part.
Atom. Nucl. 23:1 (1992), 174-238 [in Russian].
\item"[52]" Manin Yu.I., Critical dimensions of string theories and dualizing
sheaf on the (super)cur\-ves moduli space. Funkt. anal. i ego prilozh. 20:3
(1986), 88-89 [in Russian].
\item"[53]" Friedan D., Shenker S., The analytic geometry of two--dimensional
conformal field theory. Nucl. Phys. B281 (1987), 509-545.
\item"[54]" Kontsevich M.L., Virasoro algebra and Teichm\"uller spaces. Funkt.
anal. i ego prilozh. 21:2 (1987), 78-79 [in Russian].
\item"[55]" Arbarello A., De Concini C., Kac V., Procesi C., Moduli spaces of
curves and representation theory. Commun. Math. Phys. 117 (1988), 1-36.
\item"[56]" Beilinson A.A., Manin Yu.I., Schechtman V.V., Sheaves of Virasoro
and Neveu--Schwarz algebras. Lect. Notes Math. 1289 (1987), 52-66.
\item"[57]" Beilinson A.A., Schechtman V.V., Determinant bundles and Virasoro
algebra. Commun. Math. Phys. 118 (1988), 651-701.
\item"[58]" Scherk J., An introduction to the theory of dual models and
strings. Rev. Mod. Phys. 47 (1975), 123-164.
\item"[59]" Schwarz J., Dual resonance theory. Phys. Rep. C8 (1973), 269-335.
\item"[60]" Polyakov A.M., Quantum geometry of bosonic string. Phys. Lett.
B103 (1981), 207-210.
\item"[61]" Banks T., Peskin M., Gauge invariance of string fields. Nucl.
Phys. B264 (1986), 513-547.
\item"[62]" Siegel W., String field theory via BRST. In {\it Workshop on
Unified String Theories}, Singapore, 1986, 593-606.
\item"[63]" Juriev D., Infinite dimensional geometry and closed string--filed
theory. Lett. Math. Phys. 22 (1991), 1-6.
\item"[64]" Beilinson A.A., Manin Yu.I., The Mumford form and the Polyakov
measure in string theory. Commun. Math. Phys. 107, 359-376.
\item"[65]" Morozov A.Yu., Perelomov A.M., Complex geometry and string theory.
Current Probl. Math., Modern Acievements [in Russian].
\item"[66]" Morozov A.Yu., Rosly A.A., Riemann surfaces with boundary and
string theory. ZhETP 95:2 (1989), 428-441 [in Russian].
\item"[67]" Feigin B.L., Semi--infinite homology of Kac--Moody and Virasoro
algebras. Uspekhi Ma\-tem. Nauk 39:2 (1984), 195-196 [in Russian].
\item"[68]" Frenkel I., Garland H., Zukerman G., Semi--infinite cohomology and
string theory. Proc. Nat'l Acad. Sci. USA 83 (1986), 8442.
\item"[69]" Kato M., Ogawa K., Covariant quantization of string based on BRS
invariance. Nucl. Phys. B212 (1983), 443-460.
\item"[70]" Juriev D., Cohomology of the Virasoro algebra with coefficients in
string fields. Lett. Math. Phys. 19 (1990), 355-356.
\item"[71]" Juriev D., Infinite dimensional geometry and generally covariant
closed string--field theory. Lett. Math. Phys. 22 (1991), 11-14.
\item"[72]" Juriev D., Gauge inavriance of Banks--Peskin differential forms
(flat background). Lett. Math. Phys. 19 (1990), 59-64.
\item"[73]" Kirillov A.A., Geometric quantization. Current Probl. Math.,
Fundam. Directions 4 (1985), 141-178 [in Russian].
\endroster
\endRefs
\enddocument